\documentclass[12pt]{article}
\usepackage[round]{natbib}
\usepackage[english]{babel}
\usepackage[latin1]{inputenc}
\usepackage{graphicx}
\usepackage{rotating}
\usepackage{pdflscape}
\usepackage{epsfig}
\usepackage{setspace} 
\usepackage{array}
\newcolumntype{L}[1]{>{\raggedright\let\newline\\\arraybackslash\hspace{0pt}}m{#1}}
\newcolumntype{C}[1]{>{\centering\let\newline\\\arraybackslash\hspace{0pt}}m{#1}}
\newcolumntype{R}[1]{>{\raggedleft\let\newline\\\arraybackslash\hspace{0pt}}m{#1}}

\begin{document}


\title{On quantile residuals in beta regression}






\author{Gustavo H. A. Pereira}
\date{}

\maketitle

\vspace{-11.5mm} \noindent 
\begin{center}
Department of Statistics, Federal University of São Carlos
\end{center}

\vspace{1mm}

\begin{abstract}
Beta regression is often used to model the relationship between a dependent variable that assumes values on the open interval $(0,1)$ and a set of predictor variables. An important challenge in beta regression is to find residuals whose distribution is well approximated by the standard normal distribution. Two previous works compared residuals in beta regression, but the authors did not include the quantile residual. Using Monte Carlo simulation techniques, this paper studies the behavior of certain residuals in beta regression in several scenarios. Overall, the results suggest that the distribution of the quantile residual is better approximated by the standard normal distribution than that of the other residuals in most scenarios. Three applications illustrate the effectiveness of the quantile residual.
\end{abstract}

\vspace{0mm} \noindent {\textbf{Key words}}: beta regression; diagnostic analysis;
quantile residuals.

\vspace{3mm}

\section{Introduction}
\label{sec:int}

Beta regression is often used to model the relationship between a dependent variable that assumes values on the open interval $(0,1)$ and a set of predictor variables. It is useful to model rates and proportions, quantities that are common in different areas. Beta regression was introduced by \citet{pao01}, \citet{kie+mcc03} and \citet{fer+cri04}. Recent contributions in this area were made by \citet{bay+cri13}, \citet{fig+are+fer13}, \citet{cri+que14}, \citet{car+fer+val14}, \citet{esp+fer+cri14} and \citet{latif2015d}.

Residuals play an important role in checking model adequacy and in the identification of outliers and influential observations. An important challenge in regression models is to find residuals whose distribution is well approximated by the standard normal distribution. When the distribution of a residual is not well approximated by the standard normal distribution, it is difficult to interpret residuals plots. This is especially true because, when this happens, in general, residuals are not approximately identically distributed (see, for example, Table 1 of \citet{anh+san+bot14}). 

Several residuals were proposed to perform diagnostic analysis in beta regression. \citet{fer+cri04} suggested the use of two residuals: standardized and deviance. \citet{esp+fer+cri08a} introduced two other residuals based on Fisher's scoring iterative algorithm for estimating the parameters of the beta regression model, and named them as standardized weighted residual 1 and 2. Based on Monte Carlo simulation studies, the authors concluded that the distribution of the two residuals proposed in their paper is better approximated by the standard normal distribution than that of the standardized residual. The authors also revealed the shortcomings of the deviance residual and suggested it should not be used in the beta regression model. In addition, they performed diagnostic analysis in three applications and concluded that the standardized weighted residual 2 is a better residual than the standardized weighted residual 1.

A different approach was used by \citet{anh+san+bot14}. They obtained approximations, under suitable regularity conditions, for the means and variances of the standardized residual and of the standardized weighted residual 1 to order $O(n^{-1})$, where $n$ is the sample size. Using these approximations, they introduced new adjusted residuals based on the original residuals and their approximate moments.   
Employing Monte Carlo simulation studies, the authors concluded that the distribution of the adjusted standardized weighted residual 1 is better approximated by the standard normal distribution than that of the residuals proposed by \citet{esp+fer+cri08a}. 

The quantile residual \citep{dun+smy96} is a simple and general residual. It is usually used to perform diagnostic analysis in complex regression models, such as the generalized additive models for location scale and shape \citep{rig+sta05}. However, the properties of the quantile residual in the beta regression model have never been studied. The quantile residual is asymptotically normally distributed, but it is important to study its properties when sample size is not large.

The main goal of this paper is to study the properties of the quantile residual in the beta regression model. Using Monte Carlo simulation studies and three applications, we compare the behavior of the quantile residual, the standardized weighted residual 1 and 2 and the adjusted standardized weighted residual 1 in several scenarios. 

The remainder of the paper is organized as follows. Section 2 defines the beta regression model and the residuals considered in this paper. In the following section, Monte Carlo simulation studies are performed to compare the properties of the defined residuals. Three applications are presented in Section 4. Concluding remarks are provided in Section 5.

\section{Beta regression residuals}
\label{sec:betareg}

Let $y_1, y_2, \ldots, y_n$ be independent random variables, where each $y_i$, $i=1,2,\ldots,n$, is beta distributed with parameters $\mu_i$ and $\phi$, that is, each $y_i$ has density 
\begin{equation}
\label{eq:bereg}
f(y;\mu_i,\phi)=\frac{\Gamma(\phi)}{\Gamma(\mu_i\phi)\Gamma((1-\mu_i)\phi)}y^{\mu_i\phi-1}(1-y)^{(1-\mu_i)\phi-1}, \quad y \in (0,1),
\end{equation}
where $0 < \mu_i < 1$ and $\phi > 0$. In this parameterization of the beta distribution, $\mbox{E}(y_i)=\mu_i$ and $\mbox{Var}(y_i)=\mu_i(1-\mu_i)/(1+\phi)$ \citep{fer+cri04}. A beta regression model is defined assuming the following systematic
component $g(\mu_i)=x_{i}^\top\beta$, where $x_{i}=(x_{i1},x_{i2},\ldots,x_{ik})^\top$ is a vector of known covariates ($k < n$), $\beta=(\beta_1,\beta_2,\ldots,\beta_k)^\top$ is a vector of unknown parameters ($\beta \in R^k$) and $g(.)$ 
is a strictly monotonic and twice differentiable link function that maps $(0,1)$ into $R$. The parameters of the beta regression model can be estimated by maximum likelihood using a nonlinear optimization algorithm, such as a quasi-Newton algorithm \citep{noc+wri06}. The score function and the Fisher information matrix of the model are presented in \citet{fer+cri04}. The \textit{betareg} package, which is available for the R statistical software, can be used for fitting beta regression \citep{cri+zei10}.

The residuals considered in this paper are the standardized weighted residual 1 and 2 \citep{esp+fer+cri08a}, the adjusted standardized weighted residual 1 \citep{anh+san+bot14} and the quantile residual \citep{dun+smy96}. The deviance residual was not used because of its practical problems in beta regression as discussed in \citet{esp+fer+cri08a}.
In the following, the residuals considered in this paper are defined for beta regression. Here, $\hat{\beta}$ and $\hat{\phi}$ are the maximum likelihood estimators of $\beta$ and $\phi$, respectively, and $\hat{\mu}_i=g^{-1}(x_i^\top\hat{\beta})$.

The standardized weighted residual 1 is defined as 
\begin{equation}
\label{eq:rewei1}
r_i^\ast= \frac{y_i^\ast - \hat{\mu}_i^\ast}{\sqrt{\hat{\upsilon}_i^\ast}},
\end{equation}
where $y_i^\ast=\log\{y_i/(1-y_i)\}$, $\hat{\mu}_i^\ast=\widehat{\mbox{E}}(y_i^\ast)=\psi(\hat{\mu}_i\hat{\phi})-\psi((1-\hat{\mu}_i)\hat{\phi})$, $\hat{\upsilon}_i^\ast=\widehat{\mbox{Var}}(y_i^\ast)=\psi'(\hat{\mu}_i\hat{\phi})+\psi'((1-\hat{\mu}_i)\hat{\phi})$ and $\psi(.)$ is the digamma function. 

The standardized weighted residual 2 is given by
\begin{equation}
\label{eq:rewei2}
r_i^{\ast\ast}= \frac{y_i^\ast - \hat{\mu}_i^\ast}{\sqrt{\hat{\upsilon}_i^\ast(1-h_{ii})}},
\end{equation}
where $h_{ii}$ is the $i$th diagonal element of $H=\widehat{W}^{1/2}X(X^\top\widehat{W}X)^{1}X^\top\widehat{W}^{1/2}$, $X=(x_1,x_2,\ldots,x_n)^\top$ and $\widehat{W}=\mbox{diag}\{\hat{w}_1,\hat{w}_1\ldots,\hat{w}_n\}$, $\hat{w}_i=\hat{\phi}\hat{\upsilon}_i^\ast[1/\{g'(\hat{\mu}_i)\}^2]$. 

The adjusted standardized weighted residual 1 was derived by \citet{anh+san+bot14} only for the logit link function case, that is,  $g(\mu_i)=\log\{\mu_i/(1-\mu_i)\}$. It is defined as 
\begin{equation}
\label{eq:readwei1}
r_i^{\ast^a}=\frac{r_i^\ast - \widehat{\mbox{E}}(r_i^\ast)}{\widehat{\mbox{Var}}(r_i^\ast)},
\end{equation}
where $r_i^\ast$ is defined in (\ref{eq:rewei1}) and $\widehat{\mbox{E}}(r_i^\ast)$ and $\widehat{\mbox{Var}}(r_i^\ast)$ are the $i$th term of the vectors $r^\ast$ and $s^\ast$, respectively. Each of these vectors are functions of 11 vectors and matrices and their formulas can be seen in \citet{anh+san+bot14}. 

The quantile residual is given by
\begin{equation}
\label{eq:readwei1}
r_i^q = \Phi^{-1}\{F(y_i;\hat{\mu}_i,\hat{\phi})\}            
\end{equation}
where $\Phi(.)$ and $F(.)$ are the cumulative distribution function of the standard normal distribution and 
of the beta distribution, respectively.

\section{Simulation studies}
\label{sec:simulation}

Monte Carlo simulation studies were performed using a beta regression model, in which,
\begin{equation}
\label{eq:modsim}
\log\left\{\frac{\mu_i}{1-\mu_i}\right\}=\beta_1 + \beta_2 x_{i2} + \beta_3 x_{i3}, \quad i=1,2,\ldots,n.
\end{equation}
Five scenarios were considered. The first four scenarios are those used by \citet{anh+san+bot14}. The fifth scenario was inspired by real data. The median of the observed values of the response variable is on the interval $(0.15,0.30)$  in certain databases used in some beta regression papers, such as Praterï¿½s gasoline data and food expenditure data \citep{fer+cri04}. For this reason, in the fifth scenario the median of $\mu_1,\mu_2,\ldots,\mu_n$ is on the interval $(0.15, 0.30)$. For the first three scenarios, the covariates values were generated as independent draws from the standard uniform distribution. For the other scenarios, the $x_{i2}$ values were generated from the exponential distribution with mean equal to 2 and the $x_{i3}$ values from the standard normal distribution. The covariates values remained constant throughout the simulations. In the first scenario (Scenario \textrm{I}), $\beta_1=-2.3$, $\beta_2=-1.1$ and $\beta_3=-0.7$, which resulted in mean response values close to zero, $\mu \in (0.016, 0.091)$. In the second scenario (Scenario \textrm{II}), $\beta_1=-0.3$, $\beta_2=0.3$ and $\beta_3=0.7$, which resulted in mean response values close to $0.5$, $\mu \in (0.425, 0.668)$. In the following scenario (Scenario \textrm{III}), $\beta_1=4.0$, $\beta_2=-0.3$ and $\beta_3=-0.5$, which resulted in mean response values close to one, $\mu \in (0.960, 0.982)$. In the fourth scenario (Scenario \textrm{IV}), $\beta_1=1.0$, $\beta_2=0.5$ and $\beta_3=-0.5$, which yielded $\mu \in (0.608, 0.924)$. In the last scenario (Scenario \textrm{V}), $\beta_1=-2.5$, $\beta_2=2.0$ and $\beta_3=-0.5$, which yielded $\mu \in (0.057, 0.687)$. For each scenario, two different values of $\phi$ were considered: $10$ and $100$. All results are based on $5000$ Monte Carlo replications and $n = 16$. Simulations were performed using the Ox language \citep{dor07}.

Tables \ref{ta:cen1_10} to \ref{ta:cen5_10} present the sample mean, variance, skewness and kurtosis coefficients for $r_i^\ast$, $r_i^{\ast^a}$ and $r_i^q$ out of $5000$ values for $\phi = 10$ and Scenarios \textrm{I}, \textrm{II}, \textrm{IV} and \textrm{V}. Table for Scenario \textrm{III} was omitted because their results are similar to Scenario \textrm{I}. Results for $r_i^{\ast\ast}$ were also omitted from the tables.  As already noted by \citet{anh+san+bot14} and confirmed in our simulations, the distribution of $r_i^{\ast\ast}$ is worse approximated by the standard normal distribution than that of $r_i^{\ast}$ in all scenarios. 
Residual $r_i^{\ast\ast}$ has mean, skewness and kurtosis coefficients similar to $r_i^{\ast}$, but its variance is far from $1$ than that of $r_i^{\ast}$. 
The tables also contain the value of the Anderson-Darling statistic \citep{ste86} used to test whether each residual is standard normally distributed. The critical value for this Anderson-Darling test is $2.492$ at the $5\%$ significance level. However, as each test uses $5000$ replications, it is expected that most of them reject the normality hyphotesis. Therefore, the value of the Anderson-Darling statistic is used as a closeness measure between each residual distribution and standard normal distribution. 

In all scenarios, the mean of the three residuals is close to zero. In scenarios \textrm{II}, \textrm{IV} and \textrm{V}, the variance of $r_i^{\ast^a}$ is closer to one than that of the other two residuals, but none of the residuals have variance very close to one for all $16$ observations in scenarios \textrm{I} and \textrm{III}. The skewness coefficient is close to zero in all scenarios for $r_i^q$  and only in scenario \textrm{II} for $r_i^{\ast}$ and $r_i^{\ast^a}$. In scenarios \textrm{IV} and \textrm{V}, the skewness coefficient of $r_i^{\ast}$ and $r_i^{\ast^a}$ is not far from zero, but in scenarios \textrm{I} and \textrm{III} these residuals are considerably skewed. The kurtosis coefficient is not very close to $3$ in any of the scenarios for the three residuals. However, only in scenarios \textrm{I} and \textrm{III} and residuals $r_i^{\ast}$ and $r_i^{\ast^a}$, the kurtosis coefficient is extremely far from $3$ for some observations. In scenarios \textrm{I} and \textrm{III}, the Anderson-Darling statistic (ADS) is considerably smaller for $r_i^q$ than that for $r_i^{\ast^a}$ for all $16$ observations. On the other hand, in scenario \textrm{II}, the ADS is smaller for $r_i^{\ast^a}$ than that for $r_i^q$ for $15$ out of $16$ observations. In scenarios \textrm{IV} and \textrm{V}, the sample mean and the sample standard deviation of the ADS is smaller for $r_i^q$ than that for $r_i^{\ast^a}$, but the latter has a lower ADS for several observations. Residual $r_i^{\ast}$ has a larger ADS than $r_i^q$ and $r_i^{\ast^a}$ in scenarios \textrm{IV} and \textrm{V} but a similar ADS than $r_i^{\ast^a}$ in scenarios \textrm{I} and \textrm{III} and a similar ADS than $r_i^q$ in scenario \textrm{II}.

     \begin{table} [!hpb]    \caption {Simulation results for Scenario \textrm{I} and $\phi=10$.} 
    \vspace{-0.3cm} 
    \begin{center}
\tabcolsep=0.1cm    
\scalebox{0.8}{
\begin{tabular} {rrrrrrrrrrrrrrrrrrrrr}      
    \hline
 i & $\mu_i$ & \multicolumn{3} {c} {Mean} & & \multicolumn{3} {c} {Variance} & & \multicolumn{3} {c} {Skewness} & & \multicolumn{3} {c} {Kurtosis} & & \multicolumn{3} {c} {A-D Statistic} \\
\cline{3-5} \cline{7-9} \cline{11-13} \cline{15-17} \cline{19-21}
& & $r_i^{\ast}$ & $r_i^{\ast^a}$ & $r_i^q$ & & $r_i^{\ast}$ & $r_i^{\ast^a}$ & $r_i^q$ & & $r_i^{\ast}$ & $r_i^{\ast^a}$ & $r_i^q$ & & $r_i^{\ast}$ & $r_i^{\ast^a}$ & $r_i^q$ & & $r_i^{\ast}$ & $r_i^{\ast^a}$ & $r_i^q$ \\ 
\hline
1	&	$0.052$	&	$0.00$	&	$0.01$	&	$0.01$	& &	$1.01$	&	$1.03$	&	$1.10$	& &	$-0.82$	&	$-0.82$	&	$0.08$	& &	$3.43$	&	$3.44$	&	$2.38$	& &	$57.4$	&	$63.0$	&	$17.0$	\\
2	&	$0.017$	&	$0.06$	&	$0.00$	&	$0.01$	& &	$0.65$	&	$1.71$	&	$0.86$	& &	$-0.96$	&	$-0.99$	&	$0.32$	& &	$3.79$	&	$3.93$	&	$2.72$	& &	$151.8$	&	$199.2$	&	$12.6$	\\
3	&	$0.028$	&	$-0.02$	&	$0.01$	&	$-0.01$	& &	$0.98$	&	$0.98$	&	$1.05$	& &	$-1.22$	&	$-1.22$	&	$0.06$	& &	$4.69$	&	$4.70$	&	$2.54$	& &	$104.4$	&	$116.9$	&	$8.0$	\\
4	&	$0.030$	&	$-0.04$	&	$-0.01$	&	$-0.02$	& &	$1.06$	&	$1.01$	&	$1.12$	& &	$-1.13$	&	$-1.13$	&	$0.07$	& &	$4.17$	&	$4.18$	&	$2.49$	& &	$99.1$	&	$110.3$	&	$16.0$	\\
5	&	$0.066$	&	$0.02$	&	$0.01$	&	$0.00$	& &	$0.83$	&	$1.04$	&	$0.92$	& &	$-0.62$	&	$-0.63$	&	$0.10$	& &	$2.98$	&	$2.99$	&	$2.46$	& &	$51.0$	&	$46.1$	&	$3.6$	\\
6	&	$0.041$	&	$-0.02$	&	$-0.02$	&	$-0.03$	& &	$0.93$	&	$1.01$	&	$1.01$	& &	$-0.92$	&	$-0.93$	&	$0.13$	& &	$3.78$	&	$3.81$	&	$2.54$	& &	$63.8$	&	$65.2$	&	$8.7$	\\
7	&	$0.039$	&	$0.04$	&	$-0.01$	&	$0.00$	& &	$0.65$	&	$1.41$	&	$0.81$	& &	$-0.66$	&	$-0.69$	&	$0.27$	& &	$3.07$	&	$3.24$	&	$2.61$	& &	$96.1$	&	$96.5$	&	$15.7$	\\
8	&	$0.076$	&	$0.01$	&	$-0.02$	&	$-0.02$	& &	$0.70$	&	$1.15$	&	$0.80$	& &	$-0.44$	&	$-0.45$	&	$0.22$	& &	$2.83$	&	$2.90$	&	$2.51$	& &	$48.7$	&	$27.3$	&	$17.9$	\\
9	&	$0.045$	&	$-0.03$	&	$0.00$	&	$0.00$	& &	$1.12$	&	$1.03$	&	$1.17$	& &	$-1.01$	&	$-1.01$	&	$0.04$	& &	$4.04$	&	$4.05$	&	$2.43$	& &	$75.6$	&	$81.3$	&	$24.2$	\\
10	&	$0.037$	&	$-0.01$	&	$-0.01$	&	$-0.03$	& &	$0.84$	&	$1.02$	&	$0.96$	& &	$-0.86$	&	$-0.87$	&	$0.18$	& &	$3.59$	&	$3.59$	&	$2.55$	& &	$68.1$	&	$62.0$	&	$8.6$	\\
11	&	$0.036$	&	$-0.03$	&	$0.00$	&	$0.00$	& &	$1.12$	&	$1.04$	&	$1.15$	& &	$-1.25$	&	$-1.25$	&	$0.03$	& &	$4.98$	&	$4.98$	&	$2.58$	& &	$92.6$	&	$105.4$	&	$15.3$	\\
12	&	$0.033$	&	$-0.04$	&	$0.00$	&	$-0.02$	& &	$1.09$	&	$1.00$	&	$1.10$	& &	$-1.32$	&	$-1.32$	&	$0.02$	& &	$5.27$	&	$5.27$	&	$2.60$	& &	$96.2$	&	$112.8$	&	$10.7$	\\
13	&	$0.037$	&	$0.01$	&	$-0.01$	&	$-0.01$	& &	$0.75$	&	$1.12$	&	$0.89$	& &	$-0.78$	&	$-0.80$	&	$0.19$	& &	$3.39$	&	$3.45$	&	$2.51$	& &	$77.7$	&	$61.5$	&	$9.0$	\\
14	&	$0.023$	&	$-0.01$	&	$0.00$	&	$-0.03$	& &	$0.80$	&	$1.01$	&	$0.95$	& &	$-1.02$	&	$-1.03$	&	$0.23$	& &	$3.97$	&	$4.02$	&	$2.65$	& &	$98.2$	&	$85.9$	&	$10.9$	\\
15	&	$0.061$	&	$0.01$	&	$0.01$	&	$0.01$	& &	$0.89$	&	$1.01$	&	$0.99$	& &	$-0.68$	&	$-0.68$	&	$0.12$	& &	$3.19$	&	$3.19$	&	$2.46$	& &	$45.5$	&	$44.0$	&	$5.4$	\\
16	&	$0.026$	&	$0.03$	&	$0.02$	&	$0.01$	& &	$0.76$	&	$1.08$	&	$0.94$	& &	$-0.86$	&	$-0.87$	&	$0.24$	& &	$3.38$	&	$3.39$	&	$2.52$	& &	$101.3$	&	$89.6$	&	$7.9$	\\
\hline																																	
Mean	&		&	$0.00$	&	$0.00$	&	$-0.01$	& &	$0.89$	&	$1.10$	&	$0.99$	& &	$-0.91$	&	$-0.92$	&	$0.14$	& &	$3.79$	&	$3.82$	&	$2.53$	& &	$83.0$	&	$85.4$	&	$12.0$	\\
SD	&		&	$0.03$	&	$0.01$	&	$0.01$	& &	$0.16$	&	$0.19$	&	$0.12$	& &	$0.25$	&	$0.24$	&	$0.09$	& &	$0.71$	&	$0.69$	&	$0.09$	& &	$27.5$	&	$40.5$	&	$5.4$	\\
    \hline
    \end{tabular}
    }
    \end{center}
    \label{ta:cen1_10}

    \caption {Simulation results for Scenario \textrm{II} and $\phi=10$.}
    \vspace{-0.3cm} 
    \begin{center}
\tabcolsep=0.1cm    
\scalebox{0.8}{
\begin{tabular} {rrrrrrrrrrrrrrrrrrrrr}      
    \hline
 i & $\mu_i$ & \multicolumn{3} {c} {Mean} & & \multicolumn{3} {c} {Variance} & & \multicolumn{3} {c} {Skewness} & & \multicolumn{3} {c} {Kurtosis} & & \multicolumn{3} {c} {A-D Statistic} \\
\cline{3-5} \cline{7-9} \cline{11-13} \cline{15-17} \cline{19-21}
& & $r_i^{\ast}$ & $r_i^{\ast^a}$ & $r_i^q$ & & $r_i^{\ast}$ & $r_i^{\ast^a}$ & $r_i^q$ & & $r_i^{\ast}$ & $r_i^{\ast^a}$ & $r_i^q$ & & $r_i^{\ast}$ & $r_i^{\ast^a}$ & $r_i^q$ & & $r_i^{\ast}$ & $r_i^{\ast^a}$ & $r_i^q$ \\ 
\hline
1	&	$0.535$	&	$0.00$	&	$0.00$	&	$0.00$	& &	$1.07$	&	$1.02$	&	$1.07$	& &	$0.07$	&	$0.07$	&	$0.03$	& &	$2.72$	&	$2.72$	&	$2.61$	& &	$4.5$	&	$2.0$	&	$6.5$	\\
2	&	$0.661$	&	$0.00$	&	$0.00$	&	$0.01$	& &	$0.78$	&	$0.94$	&	$0.80$	& &	$0.08$	&	$0.08$	&	$-0.07$	& &	$2.68$	&	$2.69$	&	$2.61$	& &	$14.9$	&	$1.4$	&	$11.6$	\\
3	&	$0.598$	&	$-0.01$	&	$-0.01$	&	$-0.01$	& &	$1.10$	&	$1.01$	&	$1.10$	& &	$0.15$	&	$0.15$	&	$0.04$	& &	$2.74$	&	$2.74$	&	$2.60$	& &	$8.3$	&	$3.6$	&	$9.2$	\\
4	&	$0.588$	&	$-0.01$	&	$-0.01$	&	$-0.01$	& &	$1.12$	&	$1.01$	&	$1.12$	& &	$0.13$	&	$0.13$	&	$0.03$	& &	$2.75$	&	$2.75$	&	$2.62$	& &	$10.6$	&	$3.8$	&	$11.7$	\\
5	&	$0.474$	&	$0.00$	&	$0.01$	&	$0.00$	& &	$0.94$	&	$0.99$	&	$0.95$	& &	$-0.02$	&	$-0.03$	&	$0.00$	& &	$2.61$	&	$2.61$	&	$2.53$	& &	$1.5$	&	$2.1$	&	$2.1$	\\
6	&	$0.575$	&	$0.03$	&	$0.03$	&	$0.03$	& &	$1.03$	&	$1.00$	&	$1.03$	& &	$0.06$	&	$0.06$	&	$-0.02$	& &	$2.71$	&	$2.71$	&	$2.60$	& &	$4.6$	&	$3.4$	&	$6.8$	\\
7	&	$0.601$	&	$-0.02$	&	$-0.02$	&	$-0.02$	& &	$0.78$	&	$0.95$	&	$0.80$	& &	$0.06$	&	$0.06$	&	$-0.03$	& &	$2.67$	&	$2.67$	&	$2.58$	& &	$15.5$	&	$2.8$	&	$11.5$	\\
8	&	$0.455$	&	$-0.01$	&	$-0.01$	&	$-0.01$	& &	$0.84$	&	$0.99$	&	$0.85$	& &	$-0.04$	&	$-0.04$	&	$0.00$	& &	$2.66$	&	$2.66$	&	$2.59$	& &	$7.0$	&	$1.6$	&	$5.9$	\\
9	&	$0.544$	&	$-0.01$	&	$-0.01$	&	$-0.01$	& &	$1.13$	&	$1.03$	&	$1.14$	& &	$0.05$	&	$0.05$	&	$0.00$	& &	$2.66$	&	$2.66$	&	$2.56$	& &	$11.6$	&	$3.0$	&	$14.3$	\\
10	&	$0.527$	&	$-0.01$	&	$-0.01$	&	$-0.01$	& &	$0.96$	&	$0.98$	&	$0.97$	& &	$0.00$	&	$0.00$	&	$-0.02$	& &	$2.67$	&	$2.67$	&	$2.58$	& &	$1.3$	&	$1.5$	&	$2.3$	\\
11	&	$0.575$	&	$0.01$	&	$0.01$	&	$0.01$	& &	$1.13$	&	$1.02$	&	$1.13$	& &	$0.08$	&	$0.08$	&	$-0.01$	& &	$2.72$	&	$2.72$	&	$2.60$	& &	$9.8$	&	$2.0$	&	$12.5$	\\
12	&	$0.576$	&	$0.01$	&	$0.01$	&	$0.01$	& &	$1.19$	&	$1.06$	&	$1.19$	& &	$0.06$	&	$0.06$	&	$-0.02$	& &	$2.65$	&	$2.65$	&	$2.54$	& &	$18.9$	&	$5.4$	&	$22.4$	\\
13	&	$0.518$	&	$0.01$	&	$0.02$	&	$0.01$	& &	$0.88$	&	$0.97$	&	$0.89$	& &	$0.00$	&	$0.00$	&	$-0.02$	& &	$2.68$	&	$2.68$	&	$2.60$	& &	$3.9$	&	$1.8$	&	$3.6$	\\
14	&	$0.600$	&	$0.01$	&	$0.01$	&	$0.01$	& &	$1.01$	&	$1.01$	&	$1.02$	& &	$0.02$	&	$0.02$	&	$-0.07$	& &	$2.67$	&	$2.67$	&	$2.57$	& &	$1.9$	&	$2.0$	&	$4.2$	\\
15	&	$0.515$	&	$0.00$	&	$0.00$	&	$0.00$	& &	$0.97$	&	$0.98$	&	$0.98$	& &	$-0.03$	&	$-0.03$	&	$-0.04$	& &	$2.73$	&	$2.73$	&	$2.63$	& &	$0.4$	&	$0.4$	&	$1.0$	\\
16	&	$0.575$	&	$-0.02$	&	$-0.02$	&	$-0.02$	& &	$0.96$	&	$1.01$	&	$0.97$	& &	$0.04$	&	$0.04$	&	$-0.02$	& &	$2.69$	&	$2.69$	&	$2.61$	& &	$2.5$	&	$3.3$	&	$2.1$	\\
\hline																																	
Mean	&		&	$0.00$	&	$0.00$	&	$0.00$	& &	$0.99$	&	$1.00$	&	$1.00$	& &	$0.04$	&	$0.04$	&	$-0.01$	& &	$2.69$	&	$2.69$	&	$2.59$	& &	$7.3$	&	$2.5$	&	$8.0$	\\
SD	&		&	$0.01$	&	$0.01$	&	$0.01$	& &	$0.13$	&	$0.03$	&	$0.12$	& &	$0.05$	&	$0.05$	&	$0.03$	& &	$0.04$	&	$0.04$	&	$0.03$	& &	$5.7$	&	$1.2$	&	$5.8$	\\
    \hline
    \end{tabular}
    }
    \end{center}
    \label{ta:cen2_10}
    \end{table}

     \begin{table} [!hpb]
    \caption {Simulation results for Scenario \textrm{IV} and $\phi=10$.}
    \vspace{-0.3cm} 
    \begin{center}
\tabcolsep=0.1cm    
\scalebox{0.8}{
\begin{tabular} {rrrrrrrrrrrrrrrrrrrrr}      
    \hline
 i & $\mu_i$ & \multicolumn{3} {c} {Mean} & & \multicolumn{3} {c} {Variance} & & \multicolumn{3} {c} {Skewness} & & \multicolumn{3} {c} {Kurtosis} & & \multicolumn{3} {c} {A-D Statistic} \\
\cline{3-5} \cline{7-9} \cline{11-13} \cline{15-17} \cline{19-21}
& & $r_i^{\ast}$ & $r_i^{\ast^a}$ & $r_i^q$ & & $r_i^{\ast}$ & $r_i^{\ast^a}$ & $r_i^q$ & & $r_i^{\ast}$ & $r_i^{\ast^a}$ & $r_i^q$ & & $r_i^{\ast}$ & $r_i^{\ast^a}$ & $r_i^q$ & & $r_i^{\ast}$ & $r_i^{\ast^a}$ & $r_i^q$ \\ 
\hline
1	&	$0.798$	&	$0.00$	&	$0.00$	&	$-0.01$	& &	$1.07$	&	$1.02$	&	$1.09$	& &	$0.33$	&	$0.33$	&	$-0.04$	& &	$2.78$	&	$2.78$	&	$2.52$	& &	$13.6$	&	$11.2$	&	$10.3$	\\
2	&	$0.682$	&	$0.03$	&	$0.03$	&	$0.03$	& &	$1.01$	&	$1.01$	&	$1.02$	& &	$0.19$	&	$0.19$	&	$-0.01$	& &	$2.70$	&	$2.70$	&	$2.56$	& &	$4.7$	&	$4.6$	&	$6.0$	\\
3	&	$0.608$	&	$-0.01$	&	$-0.01$	&	$-0.01$	& &	$0.98$	&	$0.98$	&	$0.99$	& &	$0.08$	&	$0.08$	&	$-0.02$	& &	$2.73$	&	$2.73$	&	$2.62$	& &	$1.6$	&	$1.6$	&	$1.6$	\\
4	&	$0.641$	&	$0.01$	&	$0.01$	&	$0.00$	& &	$1.07$	&	$1.03$	&	$1.08$	& &	$0.12$	&	$0.12$	&	$-0.03$	& &	$2.64$	&	$2.64$	&	$2.54$	& &	$6.9$	&	$4.0$	&	$8.7$	\\
5	&	$0.861$	&	$0.01$	&	$0.00$	&	$0.01$	& &	$1.02$	&	$1.00$	&	$1.04$	& &	$0.50$	&	$0.50$	&	$-0.04$	& &	$3.03$	&	$3.03$	&	$2.58$	& &	$18.4$	&	$19.6$	&	$4.8$	\\
6	&	$0.727$	&	$-0.02$	&	$-0.02$	&	$-0.02$	& &	$1.04$	&	$0.95$	&	$1.05$	& &	$0.25$	&	$0.25$	&	$-0.01$	& &	$2.86$	&	$2.86$	&	$2.66$	& &	$8.1$	&	$6.9$	&	$4.5$	\\
7	&	$0.769$	&	$0.02$	&	$0.02$	&	$0.01$	& &	$1.10$	&	$1.03$	&	$1.10$	& &	$0.34$	&	$0.34$	&	$0.01$	& &	$2.80$	&	$2.80$	&	$2.54$	& &	$14.4$	&	$10.0$	&	$10.9$	\\
8	&	$0.924$	&	$-0.02$	&	$-0.02$	&	$0.01$	& &	$0.73$	&	$1.06$	&	$0.84$	& &	$0.47$	&	$0.49$	&	$-0.24$	& &	$2.96$	&	$3.02$	&	$2.64$	& &	$45.4$	&	$23.8$	&	$13.3$	\\
9	&	$0.614$	&	$-0.01$	&	$-0.01$	&	$-0.01$	& &	$0.93$	&	$0.98$	&	$0.94$	& &	$0.12$	&	$0.12$	&	$0.00$	& &	$2.65$	&	$2.64$	&	$2.55$	& &	$2.7$	&	$2.7$	&	$1.7$	\\
10	&	$0.629$	&	$0.00$	&	$0.00$	&	$0.00$	& &	$1.06$	&	$1.04$	&	$1.07$	& &	$0.11$	&	$0.11$	&	$-0.02$	& &	$2.63$	&	$2.63$	&	$2.53$	& &	$7.5$	&	$5.6$	&	$8.9$	\\
11	&	$0.863$	&	$-0.01$	&	$-0.02$	&	$-0.01$	& &	$1.01$	&	$1.00$	&	$1.05$	& &	$0.46$	&	$0.46$	&	$-0.07$	& &	$3.00$	&	$3.00$	&	$2.53$	& &	$16.8$	&	$19.0$	&	$7.3$	\\
12	&	$0.699$	&	$-0.01$	&	$-0.01$	&	$-0.01$	& &	$1.14$	&	$1.05$	&	$1.15$	& &	$0.18$	&	$0.18$	&	$-0.05$	& &	$2.75$	&	$2.75$	&	$2.59$	& &	$13.4$	&	$5.4$	&	$15.0$	\\
13	&	$0.891$	&	$0.00$	&	$-0.01$	&	$0.01$	& &	$0.84$	&	$1.02$	&	$0.90$	& &	$0.45$	&	$0.46$	&	$-0.12$	& &	$2.92$	&	$2.93$	&	$2.58$	& &	$25.7$	&	$21.1$	&	$4.3$	\\
14	&	$0.900$	&	$0.01$	&	$0.02$	&	$0.04$	& &	$0.74$	&	$1.09$	&	$0.82$	& &	$0.37$	&	$0.37$	&	$-0.21$	& &	$2.76$	&	$2.77$	&	$2.59$	& &	$32.3$	&	$17.5$	&	$18.4$	\\
15	&	$0.788$	&	$0.01$	&	$0.02$	&	$0.02$	& &	$0.79$	&	$1.00$	&	$0.81$	& &	$0.21$	&	$0.21$	&	$-0.10$	& &	$2.67$	&	$2.67$	&	$2.56$	& &	$17.1$	&	$5.5$	&	$11.8$	\\
16	&	$0.704$	&	$-0.03$	&	$-0.03$	&	$-0.03$	& &	$0.99$	&	$0.97$	&	$1.01$	& &	$0.20$	&	$0.20$	&	$-0.02$	& &	$2.75$	&	$2.75$	&	$2.58$	& &	$7.9$	&	$7.3$	&	$4.7$	\\
\hline																																	
Mean	&		&	$0.00$	&	$0.00$	&	$0.00$	& &	$0.97$	&	$1.01$	&	$1.00$	& &	$0.27$	&	$0.28$	&	$-0.06$	& &	$2.79$	&	$2.79$	&	$2.57$	& &	$14.8$	&	$10.4$	&	$8.2$	\\
SD	&		&	$0.02$	&	$0.02$	&	$0.02$	& &	$0.13$	&	$0.03$	&	$0.11$	& &	$0.14$	&	$0.15$	&	$0.07$	& &	$0.13$	&	$0.14$	&	$0.04$	& &	$11.6$	&	$7.3$	&	$4.8$	\\
    \hline
    \end{tabular}
    }
    \end{center}
    \label{ta:cen4_10}

    \caption {Simulation results for Scenario \textrm{V} and $\phi=10$.}
    \vspace{-0.3cm} 
    \begin{center}
\tabcolsep=0.1cm    
\scalebox{0.8}{
\begin{tabular} {rrrrrrrrrrrrrrrrrrrrr}      
    \hline
 i & $\mu_i$ & \multicolumn{3} {c} {Mean} & & \multicolumn{3} {c} {Variance} & & \multicolumn{3} {c} {Skewness} & & \multicolumn{3} {c} {Kurtosis} & & \multicolumn{3} {c} {A-D Statistic} \\
\cline{3-5} \cline{7-9} \cline{11-13} \cline{15-17} \cline{19-21}
& & $r_i^{\ast}$ & $r_i^{\ast^a}$ & $r_i^q$ & & $r_i^{\ast}$ & $r_i^{\ast^a}$ & $r_i^q$ & & $r_i^{\ast}$ & $r_i^{\ast^a}$ & $r_i^q$ & & $r_i^{\ast}$ & $r_i^{\ast^a}$ & $r_i^q$ & & $r_i^{\ast}$ & $r_i^{\ast^a}$ & $r_i^q$ \\ 
\hline
1	&	$0.284$	&	$0.01$	&	$0.00$	&	$0.02$	& &	$1.09$	&	$1.02$	&	$1.10$	& &	$-0.22$	&	$-0.22$	&	$0.01$	& &	$2.66$	&	$2.66$	&	$2.51$	& &	$13.9$	&	$7.7$	&	$12.8$	\\
2	&	$0.062$	&	$0.01$	&	$0.03$	&	$0.02$	& &	$0.97$	&	$0.97$	&	$1.06$	& &	$-0.79$	&	$-0.80$	&	$0.11$	& &	$3.64$	&	$3.66$	&	$2.54$	& &	$49.0$	&	$54.3$	&	$9.1$	\\
3	&	$0.057$	&	$-0.01$	&	$0.00$	&	$0.00$	& &	$0.99$	&	$1.02$	&	$1.08$	& &	$-0.77$	&	$-0.78$	&	$0.16$	& &	$3.51$	&	$3.51$	&	$2.59$	& &	$43.3$	&	$47.0$	&	$10.3$	\\
4	&	$0.069$	&	$-0.01$	&	$0.00$	&	$0.00$	& &	$1.04$	&	$1.02$	&	$1.10$	& &	$-0.80$	&	$-0.80$	&	$0.09$	& &	$3.59$	&	$3.60$	&	$2.56$	& &	$44.1$	&	$46.8$	&	$10.8$	\\
5	&	$0.384$	&	$0.00$	&	$0.00$	&	$0.00$	& &	$1.03$	&	$1.00$	&	$1.04$	& &	$-0.06$	&	$-0.06$	&	$0.06$	& &	$2.68$	&	$2.68$	&	$2.59$	& &	$2.3$	&	$1.1$	&	$4.0$	\\
6	&	$0.162$	&	$0.02$	&	$0.01$	&	$0.03$	& &	$1.11$	&	$1.02$	&	$1.12$	& &	$-0.54$	&	$-0.54$	&	$-0.06$	& &	$3.07$	&	$3.07$	&	$2.52$	& &	$33.9$	&	$25.9$	&	$16.8$	\\
7	&	$0.231$	&	$0.02$	&	$0.01$	&	$0.03$	& &	$1.10$	&	$1.02$	&	$1.11$	& &	$-0.33$	&	$-0.33$	&	$0.00$	& &	$2.86$	&	$2.86$	&	$2.58$	& &	$18.8$	&	$11.4$	&	$12.6$	\\
8	&	$0.687$	&	$-0.04$	&	$-0.01$	&	$-0.03$	& &	$0.63$	&	$0.92$	&	$0.65$	& &	$0.17$	&	$0.18$	&	$0.00$	& &	$2.62$	&	$2.62$	&	$2.55$	& &	$59.6$	&	$5.0$	&	$44.6$	\\
9	&	$0.098$	&	$0.00$	&	$-0.01$	&	$0.00$	& &	$0.99$	&	$1.03$	&	$1.04$	& &	$-0.60$	&	$-0.61$	&	$0.10$	& &	$3.28$	&	$3.30$	&	$2.58$	& &	$25.4$	&	$24.6$	&	$5.9$	\\
10	&	$0.064$	&	$-0.04$	&	$-0.02$	&	$-0.03$	& &	$1.02$	&	$1.03$	&	$1.07$	& &	$-0.82$	&	$-0.83$	&	$0.10$	& &	$3.68$	&	$3.70$	&	$2.57$	& &	$40.1$	&	$42.1$	&	$11.4$	\\
11	&	$0.248$	&	$0.03$	&	$0.03$	&	$0.04$	& &	$1.04$	&	$1.01$	&	$1.06$	& &	$-0.26$	&	$-0.26$	&	$0.02$	& &	$2.74$	&	$2.74$	&	$2.53$	& &	$15.0$	&	$11.9$	&	$10.0$	\\
12	&	$0.093$	&	$-0.01$	&	$0.00$	&	$0.01$	& &	$1.07$	&	$1.00$	&	$1.11$	& &	$-0.68$	&	$-0.69$	&	$0.05$	& &	$3.32$	&	$3.32$	&	$2.56$	& &	$38.5$	&	$38.1$	&	$11.1$	\\
13	&	$0.240$	&	$-0.01$	&	$-0.02$	&	$-0.02$	& &	$0.82$	&	$0.96$	&	$0.85$	& &	$-0.19$	&	$-0.20$	&	$0.08$	& &	$2.67$	&	$2.68$	&	$2.55$	& &	$11.0$	&	$3.4$	&	$7.4$	\\
14	&	$0.225$	&	$0.01$	&	$-0.01$	&	$-0.01$	& &	$0.69$	&	$0.95$	&	$0.72$	& &	$-0.18$	&	$-0.18$	&	$0.09$	& &	$2.61$	&	$2.62$	&	$2.50$	& &	$34.8$	&	$3.7$	&	$25.5$	\\
15	&	$0.390$	&	$0.01$	&	$0.00$	&	$0.00$	& &	$0.79$	&	$0.98$	&	$0.80$	& &	$-0.08$	&	$-0.08$	&	$0.02$	& &	$2.62$	&	$2.62$	&	$2.53$	& &	$13.7$	&	$2.0$	&	$10.7$	\\
16	&	$0.071$	&	$-0.02$	&	$-0.01$	&	$-0.01$	& &	$1.04$	&	$1.02$	&	$1.08$	& &	$-0.82$	&	$-0.82$	&	$0.07$	& &	$3.72$	&	$3.75$	&	$2.59$	& &	$41.0$	&	$43.6$	&	$8.8$	\\
\hline																																	
Mean	&		&	$0.00$	&	$0.00$	&	$0.00$	& &	$0.96$	&	$1.00$	&	$1.00$	& &	$-0.44$	&	$-0.44$	&	$0.06$	& &	$3.08$	&	$3.09$	&	$2.55$	& &	$30.3$	&	$23.0$	&	$13.2$	\\
SD	&		&	$0.02$	&	$0.01$	&	$0.02$	& &	$0.15$	&	$0.03$	&	$0.15$	& &	$0.33$	&	$0.33$	&	$0.05$	& &	$0.44$	&	$0.45$	&	$0.03$	& &	$16.3$	&	$19.4$	&	$9.7$	\\
   \hline
    \end{tabular}
    }
    \end{center}
    \label{ta:cen5_10}
    \end{table}

Tables \ref{ta:cen1_100} to \ref{ta:cen5_100} present the simulation results for $\phi = 100$. The main differences are observed in the skewness coefficient. In all scenarios, it reduces considerably when $\phi = 100$, especially for residuals $r_i^{\ast}$ and $r_i^{\ast^a}$ for which the skewness coefficients are higher when $\phi = 10$. Even in scenarios \textrm{I} and \textrm{III}, the skewness coefficient of $r_i^{\ast}$ and of $r_i^{\ast^a}$ are not far from zero when $\phi = 100$. As a consequence, except for scenario \textrm{II}, ADS decreases substantially for residuals $r_i^{\ast}$ and $r_i^{\ast^a}$ and it reduces slightly for $r_i^q$. When $\phi = 100$, in scenarios \textrm{IV} and \textrm{V}, the sample mean and the sample standard deviation of the ADS are smaller for $r_i^{\ast^a}$ than that for $r_i^q$, but the latter is still better in scenarios \textrm{I} and \textrm{III}.

     \begin{table} [!hpb]    \caption {Simulation results for Scenario \textrm{I} and $\phi=100$.}
    \vspace{-0.3cm} 
    \begin{center}
\tabcolsep=0.1cm    
\scalebox{0.8}{
\begin{tabular} {rrrrrrrrrrrrrrrrrrrrr}      
    \hline
 i & $\mu_i$ & \multicolumn{3} {c} {Mean} & & \multicolumn{3} {c} {Variance} & & \multicolumn{3} {c} {Skewness} & & \multicolumn{3} {c} {Kurtosis} & & \multicolumn{3} {c} {A-D Statistic} \\
\cline{3-5} \cline{7-9} \cline{11-13} \cline{15-17} \cline{19-21}
& & $r_i^{\ast}$ & $r_i^{\ast^a}$ & $r_i^q$ & & $r_i^{\ast}$ & $r_i^{\ast^a}$ & $r_i^q$ & & $r_i^{\ast}$ & $r_i^{\ast^a}$ & $r_i^q$ & & $r_i^{\ast}$ & $r_i^{\ast^a}$ & $r_i^q$ & & $r_i^{\ast}$ & $r_i^{\ast^a}$ & $r_i^q$ \\ 
\hline
1	&	$0.052$	&	$0.01$	&	$0.01$	&	$0.02$	& &	$1.06$	&	$1.02$	&	$1.07$	& &	$-0.26$	&	$-0.26$	&	$0.04$	& &	$2.69$	&	$2.69$	&	$2.57$	& &	$12.4$	&	$10.0$	&	$7.6$	\\
2	&	$0.017$	&	$-0.03$	&	$-0.02$	&	$-0.03$	& &	$0.94$	&	$1.03$	&	$0.96$	& &	$-0.47$	&	$-0.48$	&	$0.09$	& &	$2.98$	&	$2.98$	&	$2.59$	& &	$14.3$	&	$15.3$	&	$5.6$	\\
3	&	$0.028$	&	$-0.02$	&	$-0.01$	&	$-0.01$	& &	$1.11$	&	$1.03$	&	$1.11$	& &	$-0.46$	&	$-0.46$	&	$-0.01$	& &	$2.97$	&	$2.98$	&	$2.56$	& &	$20.1$	&	$15.6$	&	$10.9$	\\
4	&	$0.030$	&	$0.01$	&	$0.01$	&	$0.02$	& &	$1.12$	&	$1.03$	&	$1.12$	& &	$-0.49$	&	$-0.49$	&	$-0.06$	& &	$2.90$	&	$2.90$	&	$2.55$	& &	$32.4$	&	$25.6$	&	$14.9$	\\
5	&	$0.066$	&	$0.03$	&	$0.03$	&	$0.02$	& &	$0.90$	&	$0.98$	&	$0.92$	& &	$-0.16$	&	$-0.16$	&	$0.09$	& &	$2.67$	&	$2.67$	&	$2.64$	& &	$7.3$	&	$6.3$	&	$2.6$	\\
6	&	$0.041$	&	$0.01$	&	$0.00$	&	$0.01$	& &	$1.03$	&	$1.01$	&	$1.04$	& &	$-0.38$	&	$-0.38$	&	$-0.03$	& &	$2.81$	&	$2.81$	&	$2.57$	& &	$15.1$	&	$14.0$	&	$4.7$	\\
7	&	$0.039$	&	$0.00$	&	$0.00$	&	$-0.01$	& &	$0.81$	&	$1.00$	&	$0.82$	& &	$-0.26$	&	$-0.25$	&	$0.05$	& &	$2.64$	&	$2.63$	&	$2.55$	& &	$16.0$	&	$8.1$	&	$8.3$	\\
8	&	$0.076$	&	$-0.03$	&	$-0.03$	&	$-0.04$	& &	$0.69$	&	$0.93$	&	$0.70$	& &	$-0.07$	&	$-0.07$	&	$0.11$	& &	$2.56$	&	$2.55$	&	$2.57$	& &	$33.4$	&	$3.6$	&	$36.8$	\\
9	&	$0.045$	&	$0.00$	&	$0.00$	&	$0.01$	& &	$1.13$	&	$1.02$	&	$1.13$	& &	$-0.38$	&	$-0.38$	&	$-0.02$	& &	$2.90$	&	$2.90$	&	$2.65$	& &	$19.2$	&	$12.3$	&	$10.4$	\\
10	&	$0.037$	&	$0.02$	&	$0.01$	&	$0.02$	& &	$0.99$	&	$1.01$	&	$1.00$	& &	$-0.34$	&	$-0.33$	&	$0.03$	& &	$2.79$	&	$2.78$	&	$2.59$	& &	$13.7$	&	$13.9$	&	$3.9$	\\
11	&	$0.036$	&	$0.00$	&	$0.00$	&	$0.01$	& &	$1.11$	&	$1.01$	&	$1.11$	& &	$-0.40$	&	$-0.40$	&	$-0.02$	& &	$2.82$	&	$2.82$	&	$2.55$	& &	$23.1$	&	$16.1$	&	$12.8$	\\
12	&	$0.033$	&	$0.02$	&	$0.02$	&	$0.03$	& &	$1.13$	&	$1.02$	&	$1.14$	& &	$-0.43$	&	$-0.43$	&	$-0.01$	& &	$2.93$	&	$2.93$	&	$2.62$	& &	$28.6$	&	$21.0$	&	$16.0$	\\
13	&	$0.037$	&	$-0.01$	&	$-0.02$	&	$-0.02$	& &	$0.90$	&	$0.99$	&	$0.91$	& &	$-0.24$	&	$-0.24$	&	$0.11$	& &	$2.74$	&	$2.73$	&	$2.63$	& &	$6.1$	&	$4.7$	&	$4.4$	\\
14	&	$0.023$	&	$-0.01$	&	$0.00$	&	$-0.01$	& &	$1.00$	&	$0.99$	&	$1.03$	& &	$-0.40$	&	$-0.40$	&	$0.09$	& &	$2.88$	&	$2.88$	&	$2.66$	& &	$12.9$	&	$13.2$	&	$3.4$	\\
15	&	$0.061$	&	$-0.02$	&	$-0.02$	&	$-0.02$	& &	$0.93$	&	$0.96$	&	$0.94$	& &	$-0.20$	&	$-0.20$	&	$0.06$	& &	$2.69$	&	$2.69$	&	$2.63$	& &	$3.8$	&	$3.4$	&	$2.8$	\\
16	&	$0.026$	&	$0.01$	&	$0.01$	&	$0.01$	& &	$0.96$	&	$1.00$	&	$0.99$	& &	$-0.35$	&	$-0.35$	&	$0.08$	& &	$2.78$	&	$2.78$	&	$2.54$	& &	$12.0$	&	$12.7$	&	$3.5$	\\
\hline																																	
Mean	&		&	$0.00$	&	$0.00$	&	$0.00$	& &	$0.99$	&	$1.00$	&	$1.00$	& &	$-0.33$	&	$-0.33$	&	$0.04$	& &	$2.80$	&	$2.80$	&	$2.59$	& &	$16.9$	&	$12.2$	&	$9.3$	\\
SD	&		&	$0.02$	&	$0.02$	&	$0.02$	& &	$0.13$	&	$0.03$	&	$0.12$	& &	$0.12$	&	$0.12$	&	$0.06$	& &	$0.13$	&	$0.13$	&	$0.04$	& &	$8.8$	&	$6.1$	&	$8.5$	\\
   \hline
    \end{tabular}
    }
    \end{center}
    \label{ta:cen1_100}

    \caption {Simulation results for Scenario \textrm{II} and $\phi=100$.}
    \vspace{-0.3cm} 
    \begin{center}
\tabcolsep=0.1cm    
\scalebox{0.8}{
\begin{tabular} {rrrrrrrrrrrrrrrrrrrrr}      
    \hline
 i & $\mu_i$ & \multicolumn{3} {c} {Mean} & & \multicolumn{3} {c} {Variance} & & \multicolumn{3} {c} {Skewness} & & \multicolumn{3} {c} {Kurtosis} & & \multicolumn{3} {c} {A-D Statistic} \\
\cline{3-5} \cline{7-9} \cline{11-13} \cline{15-17} \cline{19-21}
& & $r_i^{\ast}$ & $r_i^{\ast^a}$ & $r_i^q$ & & $r_i^{\ast}$ & $r_i^{\ast^a}$ & $r_i^q$ & & $r_i^{\ast}$ & $r_i^{\ast^a}$ & $r_i^q$ & & $r_i^{\ast}$ & $r_i^{\ast^a}$ & $r_i^q$ & & $r_i^{\ast}$ & $r_i^{\ast^a}$ & $r_i^q$ \\ 
\hline
1	&	$0.535$	&	$0.01$	&	$0.01$	&	$0.01$	& &	$1.08$	&	$1.03$	&	$1.08$	& &	$-0.04$	&	$-0.04$	&	$-0.05$	& &	$2.64$	&	$2.64$	&	$2.64$	& &	$7.3$	&	$3.7$	&	$7.7$	\\
2	&	$0.661$	&	$0.00$	&	$0.00$	&	$0.00$	& &	$0.82$	&	$0.95$	&	$0.82$	& &	$0.10$	&	$0.10$	&	$0.06$	& &	$2.59$	&	$2.60$	&	$2.58$	& &	$9.6$	&	$2.1$	&	$8.8$	\\
3	&	$0.598$	&	$0.01$	&	$0.00$	&	$0.00$	& &	$1.09$	&	$1.01$	&	$1.09$	& &	$0.04$	&	$0.04$	&	$0.01$	& &	$2.61$	&	$2.61$	&	$2.60$	& &	$8.4$	&	$2.7$	&	$8.6$	\\
4	&	$0.588$	&	$0.01$	&	$0.01$	&	$0.01$	& &	$1.12$	&	$1.02$	&	$1.12$	& &	$0.02$	&	$0.02$	&	$-0.01$	& &	$2.58$	&	$2.58$	&	$2.56$	& &	$12.0$	&	$3.5$	&	$12.4$	\\
5	&	$0.474$	&	$0.02$	&	$0.02$	&	$0.02$	& &	$0.96$	&	$1.01$	&	$0.96$	& &	$0.02$	&	$0.02$	&	$0.03$	& &	$2.56$	&	$2.56$	&	$2.55$	& &	$2.4$	&	$3.8$	&	$2.5$	\\
6	&	$0.575$	&	$0.01$	&	$0.01$	&	$0.01$	& &	$1.02$	&	$0.99$	&	$1.02$	& &	$0.03$	&	$0.03$	&	$0.01$	& &	$2.61$	&	$2.61$	&	$2.60$	& &	$3.2$	&	$1.9$	&	$3.4$	\\
7	&	$0.601$	&	$-0.02$	&	$-0.02$	&	$-0.02$	& &	$0.82$	&	$0.97$	&	$0.82$	& &	$0.03$	&	$0.03$	&	$0.00$	& &	$2.67$	&	$2.67$	&	$2.66$	& &	$10.0$	&	$2.3$	&	$9.5$	\\
8	&	$0.455$	&	$-0.01$	&	$-0.01$	&	$-0.01$	& &	$0.83$	&	$0.96$	&	$0.83$	& &	$0.03$	&	$0.03$	&	$0.04$	& &	$2.60$	&	$2.60$	&	$2.59$	& &	$7.9$	&	$1.6$	&	$7.9$	\\
9	&	$0.544$	&	$-0.03$	&	$-0.02$	&	$-0.03$	& &	$1.13$	&	$1.03$	&	$1.13$	& &	$0.02$	&	$0.02$	&	$0.01$	& &	$2.55$	&	$2.55$	&	$2.54$	& &	$17.3$	&	$6.9$	&	$17.5$	\\
10	&	$0.527$	&	$-0.01$	&	$-0.01$	&	$-0.01$	& &	$1.00$	&	$1.00$	&	$1.00$	& &	$0.02$	&	$0.02$	&	$0.01$	& &	$2.60$	&	$2.60$	&	$2.59$	& &	$2.4$	&	$2.6$	&	$2.4$	\\
11	&	$0.575$	&	$0.00$	&	$0.00$	&	$0.00$	& &	$1.14$	&	$1.03$	&	$1.14$	& &	$0.01$	&	$0.01$	&	$-0.02$	& &	$2.60$	&	$2.60$	&	$2.59$	& &	$13.1$	&	$3.4$	&	$13.5$	\\
12	&	$0.576$	&	$-0.01$	&	$-0.01$	&	$-0.01$	& &	$1.16$	&	$1.04$	&	$1.16$	& &	$0.05$	&	$0.05$	&	$0.03$	& &	$2.57$	&	$2.57$	&	$2.56$	& &	$16.6$	&	$4.8$	&	$16.6$	\\
13	&	$0.518$	&	$-0.01$	&	$-0.01$	&	$-0.01$	& &	$0.91$	&	$0.98$	&	$0.91$	& &	$-0.03$	&	$-0.03$	&	$-0.04$	& &	$2.64$	&	$2.64$	&	$2.63$	& &	$2.0$	&	$1.3$	&	$2.0$	\\
14	&	$0.600$	&	$0.02$	&	$0.02$	&	$0.02$	& &	$0.99$	&	$0.98$	&	$0.99$	& &	$-0.02$	&	$-0.02$	&	$-0.05$	& &	$2.58$	&	$2.58$	&	$2.57$	& &	$3.0$	&	$2.9$	&	$3.6$	\\
15	&	$0.515$	&	$0.02$	&	$0.02$	&	$0.02$	& &	$0.99$	&	$0.99$	&	$0.99$	& &	$0.01$	&	$0.01$	&	$0.00$	& &	$2.61$	&	$2.61$	&	$2.60$	& &	$3.1$	&	$3.2$	&	$3.3$	\\
16	&	$0.575$	&	$0.00$	&	$0.00$	&	$0.00$	& &	$0.93$	&	$0.97$	&	$0.93$	& &	$0.05$	&	$0.05$	&	$0.03$	& &	$2.57$	&	$2.57$	&	$2.57$	& &	$1.6$	&	$1.7$	&	$1.5$	\\
\hline																																	
Mean	&		&	$0.00$	&	$0.00$	&	$0.00$	& &	$1.00$	&	$1.00$	&	$1.00$	& &	$0.02$	&	$0.02$	&	$0.00$	& &	$2.60$	&	$2.60$	&	$2.59$	& &	$7.5$	&	$3.0$	&	$7.6$	\\
SD	&		&	$0.01$	&	$0.01$	&	$0.01$	& &	$0.12$	&	$0.03$	&	$0.12$	& &	$0.03$	&	$0.03$	&	$0.03$	& &	$0.03$	&	$0.03$	&	$0.03$	& &	$5.3$	&	$1.4$	&	$5.3$	\\
   \hline
    \end{tabular}
    }
    \end{center}
    \label{ta:cen2_100}
    \end{table}

     \begin{table} [!hpb]
    \caption {Simulation results for Scenario \textrm{IV} and $\phi=100$.}
    \vspace{-0.3cm} 
    \begin{center}
\tabcolsep=0.1cm    
\scalebox{0.8}{
\begin{tabular} {rrrrrrrrrrrrrrrrrrrrr}      
    \hline
 i & $\mu_i$ & \multicolumn{3} {c} {Mean} & & \multicolumn{3} {c} {Variance} & & \multicolumn{3} {c} {Skewness} & & \multicolumn{3} {c} {Kurtosis} & & \multicolumn{3} {c} {A-D Statistic} \\
\cline{3-5} \cline{7-9} \cline{11-13} \cline{15-17} \cline{19-21}
& & $r_i^{\ast}$ & $r_i^{\ast^a}$ & $r_i^q$ & & $r_i^{\ast}$ & $r_i^{\ast^a}$ & $r_i^q$ & & $r_i^{\ast}$ & $r_i^{\ast^a}$ & $r_i^q$ & & $r_i^{\ast}$ & $r_i^{\ast^a}$ & $r_i^q$ & & $r_i^{\ast}$ & $r_i^{\ast^a}$ & $r_i^q$ \\ 
\hline
1	&	$0.798$	&	$0.00$	&	$0.00$	&	$0.00$	& &	$1.04$	&	$0.99$	&	$1.04$	& &	$0.12$	&	$0.12$	&	$0.00$	& &	$2.72$	&	$2.72$	&	$2.69$	& &	$3.4$	&	$1.7$	&	$2.6$	\\
2	&	$0.682$	&	$0.00$	&	$0.00$	&	$0.00$	& &	$1.02$	&	$1.01$	&	$1.02$	& &	$0.05$	&	$0.05$	&	$-0.01$	& &	$2.60$	&	$2.60$	&	$2.59$	& &	$3.3$	&	$3.0$	&	$3.3$	\\
3	&	$0.608$	&	$0.00$	&	$0.00$	&	$0.00$	& &	$0.99$	&	$0.99$	&	$0.99$	& &	$0.01$	&	$0.01$	&	$-0.02$	& &	$2.54$	&	$2.54$	&	$2.53$	& &	$2.8$	&	$2.8$	&	$3.0$	\\
4	&	$0.641$	&	$0.01$	&	$0.01$	&	$0.01$	& &	$1.07$	&	$1.02$	&	$1.07$	& &	$0.03$	&	$0.03$	&	$-0.01$	& &	$2.64$	&	$2.64$	&	$2.63$	& &	$5.8$	&	$2.9$	&	$6.2$	\\
5	&	$0.861$	&	$-0.02$	&	$-0.02$	&	$-0.02$	& &	$1.03$	&	$0.99$	&	$1.04$	& &	$0.13$	&	$0.13$	&	$-0.03$	& &	$2.68$	&	$2.68$	&	$2.65$	& &	$6.4$	&	$4.8$	&	$4.3$	\\
6	&	$0.727$	&	$-0.01$	&	$-0.01$	&	$-0.01$	& &	$1.11$	&	$1.02$	&	$1.11$	& &	$0.07$	&	$0.07$	&	$-0.01$	& &	$2.62$	&	$2.62$	&	$2.60$	& &	$10.6$	&	$3.5$	&	$10.3$	\\
7	&	$0.769$	&	$-0.01$	&	$-0.01$	&	$-0.01$	& &	$1.08$	&	$1.01$	&	$1.08$	& &	$0.14$	&	$0.14$	&	$0.03$	& &	$2.75$	&	$2.75$	&	$2.71$	& &	$6.6$	&	$3.0$	&	$5.1$	\\
8	&	$0.924$	&	$0.00$	&	$-0.01$	&	$0.00$	& &	$0.91$	&	$0.98$	&	$0.92$	& &	$0.14$	&	$0.14$	&	$-0.07$	& &	$2.61$	&	$2.61$	&	$2.56$	& &	$3.8$	&	$4.1$	&	$2.4$	\\
9	&	$0.614$	&	$0.00$	&	$0.00$	&	$0.00$	& &	$0.93$	&	$0.98$	&	$0.93$	& &	$0.00$	&	$0.00$	&	$-0.03$	& &	$2.57$	&	$2.57$	&	$2.57$	& &	$1.4$	&	$1.7$	&	$1.6$	\\
10	&	$0.629$	&	$0.01$	&	$0.01$	&	$0.01$	& &	$1.02$	&	$1.00$	&	$1.02$	& &	$0.03$	&	$0.03$	&	$-0.01$	& &	$2.66$	&	$2.66$	&	$2.65$	& &	$2.4$	&	$1.3$	&	$2.5$	\\
11	&	$0.863$	&	$0.02$	&	$0.02$	&	$0.01$	& &	$1.06$	&	$1.03$	&	$1.06$	& &	$0.18$	&	$0.18$	&	$0.02$	& &	$2.65$	&	$2.65$	&	$2.59$	& &	$7.5$	&	$5.7$	&	$6.0$	\\
12	&	$0.699$	&	$0.00$	&	$0.00$	&	$0.00$	& &	$1.13$	&	$1.03$	&	$1.13$	& &	$0.05$	&	$0.05$	&	$-0.01$	& &	$2.61$	&	$2.61$	&	$2.59$	& &	$12.7$	&	$4.1$	&	$13.1$	\\
13	&	$0.891$	&	$0.00$	&	$0.00$	&	$0.00$	& &	$0.91$	&	$0.98$	&	$0.92$	& &	$0.11$	&	$0.11$	&	$-0.06$	& &	$2.65$	&	$2.65$	&	$2.63$	& &	$3.2$	&	$2.6$	&	$1.9$	\\
14	&	$0.900$	&	$0.02$	&	$0.02$	&	$0.02$	& &	$0.85$	&	$0.99$	&	$0.86$	& &	$0.13$	&	$0.13$	&	$-0.05$	& &	$2.62$	&	$2.62$	&	$2.60$	& &	$6.9$	&	$3.2$	&	$6.8$	\\
15	&	$0.788$	&	$0.02$	&	$0.02$	&	$0.02$	& &	$0.79$	&	$0.96$	&	$0.80$	& &	$0.02$	&	$0.02$	&	$-0.07$	& &	$2.60$	&	$2.60$	&	$2.59$	& &	$12.7$	&	$2.5$	&	$13.9$	\\
16	&	$0.704$	&	$-0.02$	&	$-0.02$	&	$-0.02$	& &	$1.03$	&	$1.00$	&	$1.03$	& &	$0.09$	&	$0.09$	&	$0.02$	& &	$2.60$	&	$2.60$	&	$2.59$	& &	$6.2$	&	$4.9$	&	$5.3$	\\
\hline																																	
Mean	&		&	$0.00$	&	$0.00$	&	$0.00$	& &	$1.00$	&	$1.00$	&	$1.00$	& &	$0.08$	&	$0.08$	&	$-0.02$	& &	$2.63$	&	$2.63$	&	$2.61$	& &	$6.0$	&	$3.2$	&	$5.5$	\\
SD	&		&	$0.01$	&	$0.01$	&	$0.01$	& &	$0.09$	&	$0.02$	&	$0.09$	& &	$0.06$	&	$0.06$	&	$0.03$	& &	$0.05$	&	$0.05$	&	$0.05$	& &	$3.5$	&	$1.2$	&	$3.8$	\\
   \hline
    \end{tabular}
    }
    \end{center}
    \label{ta:cen4_100}

    \caption {Simulation results for Scenario \textrm{V} and $\phi=100$.}
    \vspace{-0.3cm} 
    \begin{center}
\tabcolsep=0.1cm    
\scalebox{0.8}{
\begin{tabular} {rrrrrrrrrrrrrrrrrrrrr}      
    \hline
 i & $\mu_i$ & \multicolumn{3} {c} {Mean} & & \multicolumn{3} {c} {Variance} & & \multicolumn{3} {c} {Skewness} & & \multicolumn{3} {c} {Kurtosis} & & \multicolumn{3} {c} {A-D Statistic} \\
\cline{3-5} \cline{7-9} \cline{11-13} \cline{15-17} \cline{19-21}
& & $r_i^{\ast}$ & $r_i^{\ast^a}$ & $r_i^q$ & & $r_i^{\ast}$ & $r_i^{\ast^a}$ & $r_i^q$ & & $r_i^{\ast}$ & $r_i^{\ast^a}$ & $r_i^q$ & & $r_i^{\ast}$ & $r_i^{\ast^a}$ & $r_i^q$ & & $r_i^{\ast}$ & $r_i^{\ast^a}$ & $r_i^q$ \\  
\hline
1	&	$0.284$	&	$0.02$	&	$0.02$	&	$0.02$	& &	$1.08$	&	$1.02$	&	$1.08$	& &	$-0.06$	&	$-0.06$	&	$0.01$	& &	$2.57$	&	$2.57$	&	$2.55$	& &	$10.4$	&	$5.2$	&	$10.2$	\\
2	&	$0.062$	&	$0.00$	&	$0.01$	&	$0.00$	& &	$1.11$	&	$1.03$	&	$1.12$	& &	$-0.21$	&	$-0.21$	&	$0.06$	& &	$2.64$	&	$2.63$	&	$2.56$	& &	$14.3$	&	$8.0$	&	$13.6$	\\
3	&	$0.057$	&	$-0.01$	&	$0.00$	&	$-0.01$	& &	$1.07$	&	$1.00$	&	$1.08$	& &	$-0.27$	&	$-0.27$	&	$0.02$	& &	$2.76$	&	$2.76$	&	$2.62$	& &	$9.8$	&	$6.6$	&	$6.8$	\\
4	&	$0.069$	&	$0.01$	&	$0.01$	&	$0.01$	& &	$1.07$	&	$0.99$	&	$1.07$	& &	$-0.25$	&	$-0.25$	&	$0.00$	& &	$2.66$	&	$2.67$	&	$2.56$	& &	$13.0$	&	$9.3$	&	$8.3$	\\
5	&	$0.384$	&	$0.00$	&	$0.00$	&	$0.00$	& &	$1.04$	&	$1.01$	&	$1.04$	& &	$-0.05$	&	$-0.05$	&	$-0.01$	& &	$2.58$	&	$2.58$	&	$2.57$	& &	$4.5$	&	$2.8$	&	$4.6$	\\
6	&	$0.162$	&	$0.00$	&	$-0.01$	&	$0.00$	& &	$1.11$	&	$1.03$	&	$1.11$	& &	$-0.15$	&	$-0.15$	&	$-0.01$	& &	$2.63$	&	$2.63$	&	$2.58$	& &	$11.8$	&	$4.8$	&	$10.5$	\\
7	&	$0.231$	&	$0.00$	&	$0.00$	&	$0.01$	& &	$1.11$	&	$1.04$	&	$1.11$	& &	$-0.10$	&	$-0.10$	&	$-0.01$	& &	$2.59$	&	$2.59$	&	$2.56$	& &	$12.2$	&	$5.4$	&	$12.0$	\\
8	&	$0.687$	&	$-0.02$	&	$0.00$	&	$-0.01$	& &	$0.63$	&	$0.90$	&	$0.63$	& &	$0.07$	&	$0.07$	&	$0.02$	& &	$2.65$	&	$2.64$	&	$2.64$	& &	$56.0$	&	$2.8$	&	$54.3$	\\
9	&	$0.098$	&	$-0.01$	&	$-0.01$	&	$-0.01$	& &	$1.04$	&	$1.02$	&	$1.04$	& &	$-0.21$	&	$-0.21$	&	$-0.02$	& &	$2.61$	&	$2.61$	&	$2.53$	& &	$8.6$	&	$6.9$	&	$6.4$	\\
10	&	$0.064$	&	$-0.01$	&	$-0.01$	&	$-0.01$	& &	$1.09$	&	$1.01$	&	$1.09$	& &	$-0.21$	&	$-0.21$	&	$0.06$	& &	$2.65$	&	$2.65$	&	$2.57$	& &	$11.3$	&	$6.2$	&	$10.7$	\\
11	&	$0.248$	&	$0.01$	&	$0.01$	&	$0.01$	& &	$1.04$	&	$1.01$	&	$1.04$	& &	$-0.10$	&	$-0.10$	&	$-0.02$	& &	$2.54$	&	$2.54$	&	$2.51$	& &	$6.5$	&	$4.1$	&	$6.1$	\\
12	&	$0.093$	&	$0.00$	&	$0.00$	&	$0.01$	& &	$1.10$	&	$1.01$	&	$1.11$	& &	$-0.17$	&	$-0.17$	&	$0.05$	& &	$2.70$	&	$2.70$	&	$2.64$	& &	$9.9$	&	$4.1$	&	$8.9$	\\
13	&	$0.240$	&	$0.02$	&	$0.02$	&	$0.02$	& &	$0.83$	&	$0.94$	&	$0.83$	& &	$-0.07$	&	$-0.06$	&	$0.02$	& &	$2.63$	&	$2.63$	&	$2.62$	& &	$9.1$	&	$2.6$	&	$7.6$	\\
14	&	$0.225$	&	$-0.02$	&	$-0.02$	&	$-0.02$	& &	$0.75$	&	$0.96$	&	$0.75$	& &	$-0.05$	&	$-0.05$	&	$0.03$	& &	$2.56$	&	$2.56$	&	$2.56$	& &	$19.1$	&	$3.1$	&	$19.6$	\\
15	&	$0.390$	&	$0.00$	&	$0.00$	&	$0.00$	& &	$0.77$	&	$0.95$	&	$0.77$	& &	$-0.03$	&	$-0.02$	&	$0.00$	& &	$2.52$	&	$2.52$	&	$2.51$	& &	$15.3$	&	$3.0$	&	$15.0$	\\
16	&	$0.071$	&	$-0.01$	&	$0.00$	&	$0.00$	& &	$1.10$	&	$1.02$	&	$1.10$	& &	$-0.28$	&	$-0.28$	&	$0.00$	& &	$2.81$	&	$2.81$	&	$2.67$	& &	$11.6$	&	$7.1$	&	$7.6$	\\
\hline																																	
Mean	&		&	$0.00$	&	$0.00$	&	$0.00$	& &	$1.00$	&	$1.00$	&	$1.00$	& &	$-0.13$	&	$-0.13$	&	$0.01$	& &	$2.63$	&	$2.63$	&	$2.58$	& &	$14.0$	&	$5.1$	&	$12.6$	\\
SD	&		&	$0.01$	&	$0.01$	&	$0.01$	& &	$0.16$	&	$0.04$	&	$0.16$	& &	$0.10$	&	$0.10$	&	$0.03$	& &	$0.08$	&	$0.08$	&	$0.05$	& &	$11.7$	&	$2.1$	&	$11.7$	\\
   \hline
    \end{tabular}
    }
    \end{center}
    \label{ta:cen5_100}
    \end{table}    

Simulations were also carried out for $n=40$. Table \ref{ta:comp_ad} summarizes the results of the ADS in all scenarios for $n=16$ and for $n=40$. In all scenarios, the sample mean, the sample standard deviation and the samples quartiles of the ADS reduce considerably for $r_i^q$ when sample size increases from $n=16$ to $n=40$. The reduction is smaller for $r_i^{\ast}$ and $r_i^{\ast^a}$ in some scenarios and it does not exist in others. As a consequence, except for scenario \textrm{II}, the sample mean and the sample standard deviation of the ADS are lower for $r_i^q$ than that for $r_i^{\ast}$ and $r_i^{\ast^a}$ when $n=40$. In scenario \textrm{II}, $r_i^q$ and $r_i^{\ast}$ have similar sample mean of the ADS and $r_i^{\ast^a}$ has the smallest values. 

     \begin{table} [!hpb]
    \caption {Anderson-Darling statistic for $n=16$ and $n=40$.}
    \begin{center}
\tabcolsep=0.1cm    
\scalebox{0.7}{
\begin{tabular} {crrrcrrrrrrrrrr}      
    \hline
Scenario & $\phi$ & & n & Residual & Mean & Standard & Minimum & & & & \multicolumn{3} {c} {Quartiles} & Maximum \\
\cline{12-14}
 &  &  &  & & & deviation & & & & & $Q_1$ & $Q_2$ & $Q_3$ &  \\
   \hline
1	&	10	& &	16	&	$r_i^{\ast}$	&	83.0	&	27.5	&	45.5	& & & &	62.2	&	85.1	&	98.4	&	151.8	\\
	&		& &		&	$r_i^{\ast^a}$	&	85.4	&	40.5	&	27.3	& & & &	61.9	&	83.6	&	106.6	&	199.2	\\
	&		& &		&	$r_i^q$	&	12.0	&	5.4	&	3.6	& & & &	8.4	&	10.8	&	15.8	&	24.2	\\
\hline																					
1	&	10	& &	40	&	$r_i^{\ast}$	&	105.3	&	26.2	&	55.9	& & & &	88.9	&	105.3	&	123.3	&	159.2	\\
	&		& &		&	$r_i^{\ast^a}$	&	100.9	&	28.5	&	42.1	& & & &	79.5	&	106.6	&	121.1	&	151.4	\\
	&		& &		&	$r_i^q$	&	2.7	&	2.0	&	0.5	& & & &	1.2	&	2.3	&	3.2	&	8.2	\\
\hline																					
1	&	100	& &	16	&	$r_i^{\ast}$	&	16.9	&	8.8	&	3.8	& & & &	12.3	&	14.7	&	20.9	&	33.4	\\
	&		& &		&	$r_i^{\ast^a}$	&	12.2	&	6.1	&	3.4	& & & &	7.6	&	12.9	&	15.4	&	25.6	\\
	&		& &		&	$r_i^q$	&	9.3	&	8.5	&	2.6	& & & &	3.8	&	6.6	&	11.4	&	36.8	\\
\hline																					
1	&	100	& &	40	&	$r_i^{\ast}$	&	13.1	&	5.8	&	3.3	& & & &	8.9	&	13.0	&	16.2	&	27.2	\\
	&		& &		&	$r_i^{\ast^a}$	&	12.5	&	6.1	&	2.6	& & & &	7.8	&	12.9	&	16.2	&	27.2	\\
	&		& &		&	$r_i^q$	&	1.8	&	1.3	&	0.4	& & & &	0.9	&	1.6	&	2.2	&	5.0	\\
\hline																					
2	&	10	& &	16	&	$r_i^{\ast}$	&	7.3	&	5.7	&	0.4	& & & &	2.3	&	5.8	&	10.8	&	18.9	\\
	&		& &		&	$r_i^{\ast^a}$	&	2.5	&	1.2	&	0.4	& & & &	1.7	&	2.0	&	3.3	&	5.4	\\
	&		& &		&	$r_i^q$	&	8.0	&	5.8	&	1.0	& & & &	3.3	&	6.6	&	11.6	&	22.4	\\
\hline																					
2	&	10	& &	40	&	$r_i^{\ast}$	&	2.2	&	1.9	&	0.4	& & & &	1.0	&	1.5	&	2.7	&	8.7	\\
	&		& &		&	$r_i^{\ast^a}$	&	1.5	&	1.0	&	0.3	& & & &	0.8	&	1.2	&	1.9	&	4.9	\\
	&		& &		&	$r_i^q$	&	2.0	&	1.5	&	0.2	& & & &	0.8	&	1.4	&	2.6	&	6.0	\\
\hline																					
2	&	100	& &	16	&	$r_i^{\ast}$	&	7.5	&	5.3	&	1.6	& & & &	2.9	&	7.6	&	10.5	&	17.3	\\
	&		& &		&	$r_i^{\ast^a}$	&	3.0	&	1.4	&	1.3	& & & &	2.0	&	2.8	&	3.6	&	6.9	\\
	&		& &		&	$r_i^q$	&	7.6	&	5.3	&	1.5	& & & &	3.1	&	7.8	&	10.2	&	17.5	\\
\hline																					
2	&	100	& &	40	&	$r_i^{\ast}$	&	2.0	&	1.7	&	0.3	& & & &	0.7	&	1.3	&	3.4	&	8.2	\\
	&		& &		&	$r_i^{\ast^a}$	&	1.4	&	1.1	&	0.3	& & & &	0.5	&	1.1	&	1.7	&	4.4	\\
	&		& &		&	$r_i^q$	&	1.9	&	1.7	&	0.3	& & & &	0.7	&	1.2	&	3.5	&	7.3	\\
\hline																					
4	&	10	& &	16	&	$r_i^{\ast}$	&	14.8	&	11.6	&	1.6	& & & &	7.3	&	13.5	&	17.4	&	45.4	\\
	&		& &		&	$r_i^{\ast^a}$	&	10.4	&	7.3	&	1.6	& & & &	5.2	&	7.1	&	17.9	&	23.8	\\
	&		& &		&	$r_i^q$	&	8.2	&	4.8	&	1.6	& & & &	4.6	&	8.0	&	11.1	&	18.4	\\
\hline																					
4	&	10	& &	40	&	$r_i^{\ast}$	&	14.1	&	8.1	&	1.3	& & & &	8.7	&	13.6	&	18.8	&	38.1	\\
	&		& &		&	$r_i^{\ast^a}$	&	13.1	&	8.2	&	0.8	& & & &	7.9	&	12.0	&	19.0	&	35.3	\\
	&		& &		&	$r_i^q$	&	2.0	&	1.3	&	0.4	& & & &	1.2	&	1.5	&	2.5	&	7.0	\\
\hline																					
4	&	100	& &	16	&	$r_i^{\ast}$	&	6.0	&	3.5	&	1.4	& & & &	3.3	&	6.0	&	7.0	&	12.7	\\
	&		& &		&	$r_i^{\ast^a}$	&	3.2	&	1.2	&	1.3	& & & &	2.6	&	3.0	&	4.1	&	5.7	\\
	&		& &		&	$r_i^q$	&	5.5	&	3.8	&	1.6	& & & &	2.6	&	4.7	&	6.3	&	13.9	\\
\hline																					
4	&	100	& &	40	&	$r_i^{\ast}$	&	3.3	&	2.0	&	0.9	& & & &	1.8	&	2.8	&	4.2	&	10.1	\\
	&		& &		&	$r_i^{\ast^a}$	&	2.4	&	1.5	&	0.2	& & & &	1.3	&	2.0	&	3.2	&	6.1	\\
	&		& &		&	$r_i^q$	&	2.2	&	1.6	&	0.5	& & & &	1.3	&	1.7	&	2.7	&	9.8	\\
\hline																					
5	&	10	& &	16	&	$r_i^{\ast}$	&	30.3	&	16.3	&	2.3	& & & &	14.8	&	34.3	&	41.6	&	59.6	\\
	&		& &		&	$r_i^{\ast^a}$	&	23.0	&	19.4	&	1.1	& & & &	4.6	&	18.2	&	42.5	&	54.3	\\
	&		& &		&	$r_i^q$	&	13.2	&	9.7	&	4.0	& & & &	9.1	&	10.7	&	12.6	&	44.6	\\
\hline																					
5	&	10	& &	40	&	$r_i^{\ast}$	&	26.2	&	20.3	&	3.8	& & & &	12.7	&	19.3	&	34.4	&	79.7	\\
	&		& &		&	$r_i^{\ast^a}$	&	24.9	&	20.0	&	0.8	& & & &	8.2	&	18.5	&	35.8	&	78.0	\\
	&		& &		&	$r_i^q$	&	3.0	&	2.8	&	0.4	& & & &	1.6	&	2.2	&	3.1	&	14.1	\\
\hline																					
5	&	100	& &	16	&	$r_i^{\ast}$	&	14.0	&	11.7	&	4.5	& & & &	9.6	&	11.5	&	13.3	&	56.0	\\
	&		& &		&	$r_i^{\ast^a}$	&	5.1	&	2.1	&	2.6	& & & &	3.0	&	5.0	&	6.7	&	9.3	\\
	&		& &		&	$r_i^q$	&	12.6	&	11.7	&	4.6	& & & &	7.4	&	9.6	&	12.4	&	54.3	\\
\hline																					
5	&	100	& &	40	&	$r_i^{\ast}$	&	4.4	&	3.1	&	0.5	& & & &	1.9	&	3.7	&	6.1	&	13.6	\\
	&		& &		&	$r_i^{\ast^a}$	&	3.5	&	3.0	&	0.7	& & & &	1.3	&	2.6	&	4.1	&	14.5	\\
	&		& &		&	$r_i^q$	&	2.2	&	2.2	&	0.5	& & & &	1.0	&	1.4	&	2.9	&	11.2	\\
\hline   
    \end{tabular}
    }
    \end{center}
    \label{ta:comp_ad}
    \end{table} 

The logit is the most used link function in beta regression because it facilitates interpretation of the parameters. 
However, it is a symmetric link function and this may facilitate the normal approximation of the distribution of
the residuals considered in this paper. In order to check whether the results change considerably when a asymmetrical
link function is used, the same scenarios were considered in a beta regression model with complementary log-log
link function \citep{mccullagh1989generalized}. The value of the parameters in each scenario were changed in such a way that the 
mean, standard deviation, minimum and maximum of the vector $(\mu_1,\mu_2,\ldots,\mu_n)$ keep almost the same
of the logit link function case. Table \ref{ta:cloglog} summarizes the results of the ADS. The adjusted weighted standardized 
residual 1 was not considered because it was derived by \citet{anh+san+bot14} only for the logit link function.

Except in scenario \textrm{II} where $r_i^q$ and $r_i^{\ast}$ have similar behavior, the sample mean of the ADS is considerably smaller for $r_i^q$ than that for $r_i^{\ast}$, especially when $n=40$. When the complementary log-log link function is used and $n=16$, the distribution of $r_i^q$ and $r_i^{\ast}$ is worse approximated by the standard normal distribution than that when the logit link function is used, notably in the scenarios where the predictor variables are not generated from the uniform distribution. However, $r_i^q$ have similar behavior in both cases when $n=40$. This suggests that when the complementary log-log link function is used a larger sample size is necessary, but $n=40$ is large enough to that the distribution of the $r_i^q$ can be well approximated by the standard normal distribution.

     \begin{table} [!hpb]
    \caption {Anderson-Darling statistic for the models with complementary log-log link function.}
    \begin{center}
\tabcolsep=0.1cm    
\scalebox{0.7}{
\begin{tabular} {crrrcrrrrrrrrrr}      
    \hline
Scenario & $\phi$ & & n & Residual & Mean & Standard & Minimum & & & & \multicolumn{3} {c} {Quartiles} & Maximum \\
\cline{12-14}
 &  &  &  & & & deviation & & & & & $Q_1$ & $Q_2$ & $Q_3$ &  \\
   \hline
1	&	10	& &	16	&	$R_i^{\ast}$	&	83.7	&	28.5	&	43.9	& & & &	62.9	&	86.9	&	99.6	&	158.0	\\
	&		& &		&	$R_i^q$	&	12.5	&	6.6	&	4.6	& & & &	7.2	&	11.5	&	15.4	&	28.6	\\
\hline																					
1	&	10	& &	40	&	$R_i^{\ast}$	&	103.7	&	27.1	&	56.7	& & & &	83.8	&	105.2	&	124.0	&	155.4	\\
	&		& &		&	$R_i^q$	&	2.9	&	1.6	&	0.3	& & & &	1.7	&	2.7	&	3.9	&	6.9	\\
\hline																					
1	&	100	& &	16	&	$R_i^{\ast}$	&	17.4	&	9.4	&	4.7	& & & &	11.3	&	17.9	&	19.4	&	37.7	\\
	&		& &		&	$R_i^q$	&	10.4	&	10.0	&	2.0	& & & &	3.3	&	8.5	&	13.3	&	42.5	\\
\hline																					
1	&	100	& &	40	&	$R_i^{\ast}$	&	13.0	&	5.6	&	4.2	& & & &	8.8	&	12.2	&	16.2	&	23.6	\\
	&		& &		&	$R_i^q$	&	1.8	&	1.1	&	0.2	& & & &	1.0	&	1.6	&	2.5	&	4.9	\\
\hline																					
2	&	10	& &	16	&	$R_i^{\ast}$	&	7.8	&	7.5	&	0.8	& & & &	2.8	&	4.3	&	10.8	&	26.6	\\
	&		& &		&	$R_i^q$	&	8.3	&	6.3	&	1.6	& & & &	3.0	&	6.3	&	13.1	&	20.0	\\
\hline																					
2	&	10	& &	40	&	$R_i^{\ast}$	&	2.0	&	2.0	&	0.2	& & & &	0.7	&	1.4	&	2.2	&	8.5	\\
	&		& &		&	$R_i^q$	&	1.9	&	1.6	&	0.4	& & & &	0.7	&	1.4	&	2.1	&	7.4	\\
\hline																					
2	&	100	& &	16	&	$R_i^{\ast}$	&	8.9	&	6.3	&	1.5	& & & &	4.5	&	7.1	&	14.6	&	21.9	\\
	&		& &		&	$R_i^q$	&	8.9	&	6.2	&	1.6	& & & &	4.5	&	7.2	&	14.5	&	21.0	\\
\hline																					
2	&	100	& &	40	&	$R_i^{\ast}$	&	1.5	&	0.9	&	0.2	& & & &	0.6	&	1.5	&	2.1	&	3.6	\\
	&		& &		&	$R_i^q$	&	1.6	&	0.9	&	0.2	& & & &	0.7	&	1.5	&	2.1	&	3.4	\\
\hline																					
4	&	10	& &	16	&	$R_i^{\ast}$	&	21.5	&	28.5	&	2.8	& & & &	7.7	&	10.8	&	19.2	&	108.9	\\
	&		& &		&	$R_i^q$	&	13.8	&	13.8	&	4.7	& & & &	5.9	&	9.0	&	12.6	&	54.4	\\
\hline																					
4	&	10	& &	40	&	$R_i^{\ast}$	&	15.9	&	13.1	&	2.6	& & & &	7.0	&	13.8	&	17.5	&	72.7	\\
	&		& &		&	$R_i^q$	&	2.3	&	1.8	&	0.5	& & & &	0.9	&	1.5	&	3.4	&	7.6	\\
\hline																					
4	&	100	& &	16	&	$R_i^{\ast}$	&	8.1	&	5.5	&	3.4	& & & &	5.3	&	6.7	&	7.6	&	25.6	\\
	&		& &		&	$R_i^q$	&	7.2	&	3.8	&	3.2	& & & &	4.7	&	6.4	&	8.1	&	16.3	\\
\hline																					
4	&	100	& &	40	&	$R_i^{\ast}$	&	3.2	&	1.8	&	0.6	& & & &	2.1	&	3.0	&	3.8	&	8.6	\\
	&		& &		&	$R_i^q$	&	1.9	&	1.4	&	0.4	& & & &	1.1	&	1.5	&	2.3	&	6.5	\\
\hline																					
5	&	10	& &	16	&	$R_i^{\ast}$	&	37.1	&	42.4	&	7.7	& & & &	14.9	&	32.0	&	36.0	&	190.1	\\
	&		& &		&	$R_i^q$	&	21.8	&	39.2	&	4.4	& & & &	8.1	&	10.7	&	16.6	&	166.9	\\
\hline																					
5	&	10	& &	40	&	$R_i^{\ast}$	&	21.2	&	15.2	&	2.5	& & & &	11.7	&	16.9	&	23.8	&	64.6	\\
	&		& &		&	$R_i^q$	&	2.7	&	3.1	&	0.5	& & & &	1.2	&	1.9	&	2.4	&	15.4	\\
\hline																					
5	&	100	& &	16	&	$R_i^{\ast}$	&	23.4	&	43.5	&	4.7	& & & &	9.2	&	11.8	&	15.2	&	184.6	\\
	&		& &		&	$R_i^q$	&	22.1	&	43.4	&	4.9	& & & &	8.3	&	10.2	&	12.8	&	183.0	\\
\hline																					
5	&	100	& &	40	&	$R_i^{\ast}$	&	4.6	&	4.1	&	0.7	& & & &	1.8	&	3.8	&	4.9	&	23.3	\\
	&		& &		&	$R_i^q$	&	2.9	&	4.2	&	0.4	& & & &	1.2	&	1.7	&	2.7	&	23.0	\\
\hline   
    \end{tabular}
    }
    \end{center}
    \label{ta:cloglog}
    \end{table}

\section{Applications}

In this section, we present three applications, two of them based on real data and one that
employs simulated data. Applications were performed using Ox and the betareg package for the R software. 

\begin{figure}[!htbp]\centering
\includegraphics[page=1]{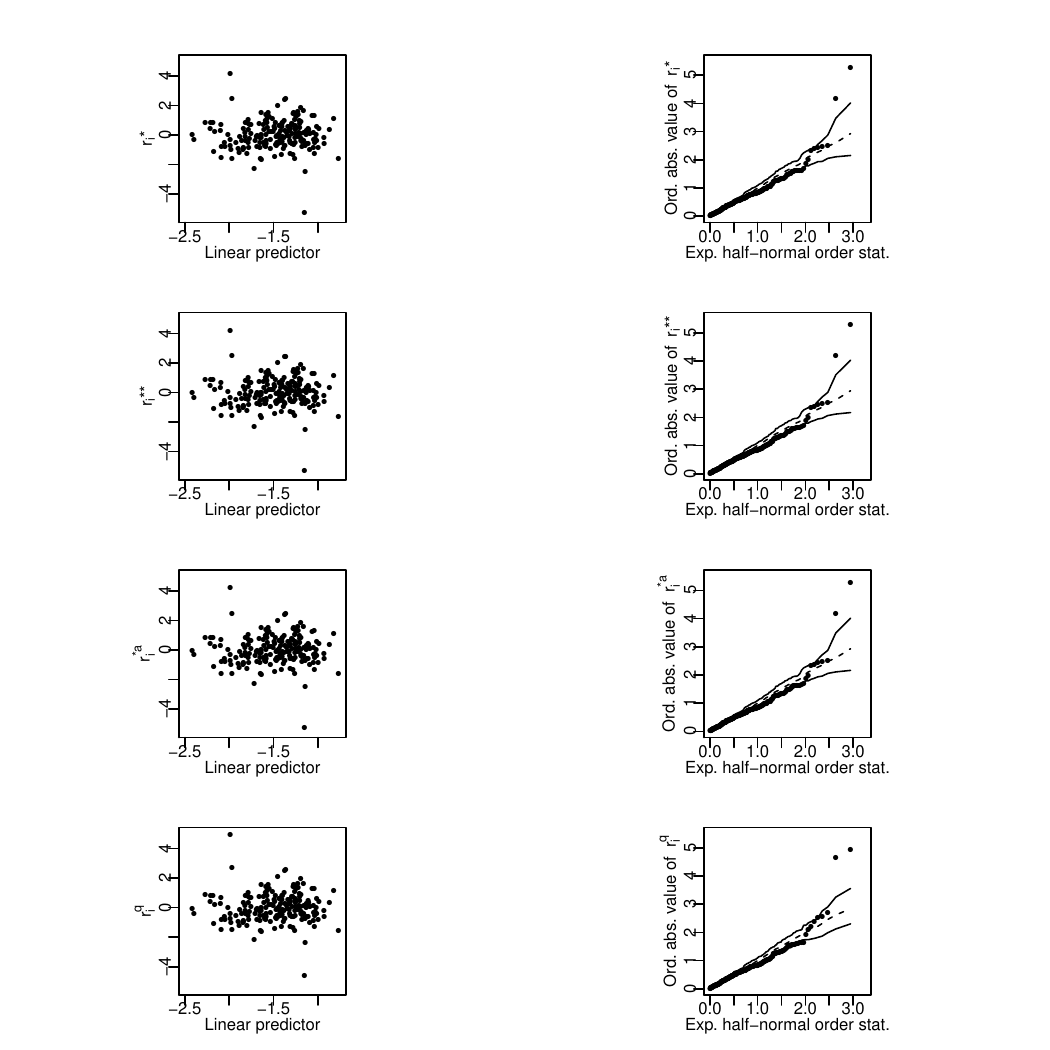}
\caption{Residuals against linear predictor (left) and half-normal residual plots with simulated envelope (right)
for Peruvian election data.}
\label{fi:env1}
\end{figure}

In Section \ref{sec:simulation}, we studied whether the distribution of the residuals considered in this paper are well approximated by the standard normal distribution. Other essential property is the ability to identify model misspecification. The first application uses data on the 2006 Peruvian general election to investigate whether $r_i^\ast$, $r_i^{\ast\ast}$, $r_i^{\ast^a}$ and $r_i^q$ have this property. The response variable is the proportion of blank votes of an electoral district and the single predictor variable is the Human Development Index.
It is known that beta regression does not provide a reasonable fit to these data, see for example \citet{bayes2012new} and \citet{lemonte2015new}, who proposed alternative models because of the lack of fit.
We fitted beta regression model with logit link function for Peruvian election data and obtained the four set of residuals. Figure \ref{fi:env1} presents a plot of residuals against linear predictor and a half-normal residual plot with simulated envelope \citep[Section 4.2]{atk85} for $r_i^\ast$, $r_i^{\ast\ast}$, $r_i^{\ast^a}$ and $r_i^q$. Clearly, the four residuals suggest lack of fit of the beta regression model, in agreement with previous findings.

\begin{figure}[!htbp]\centering
\includegraphics[page=2]{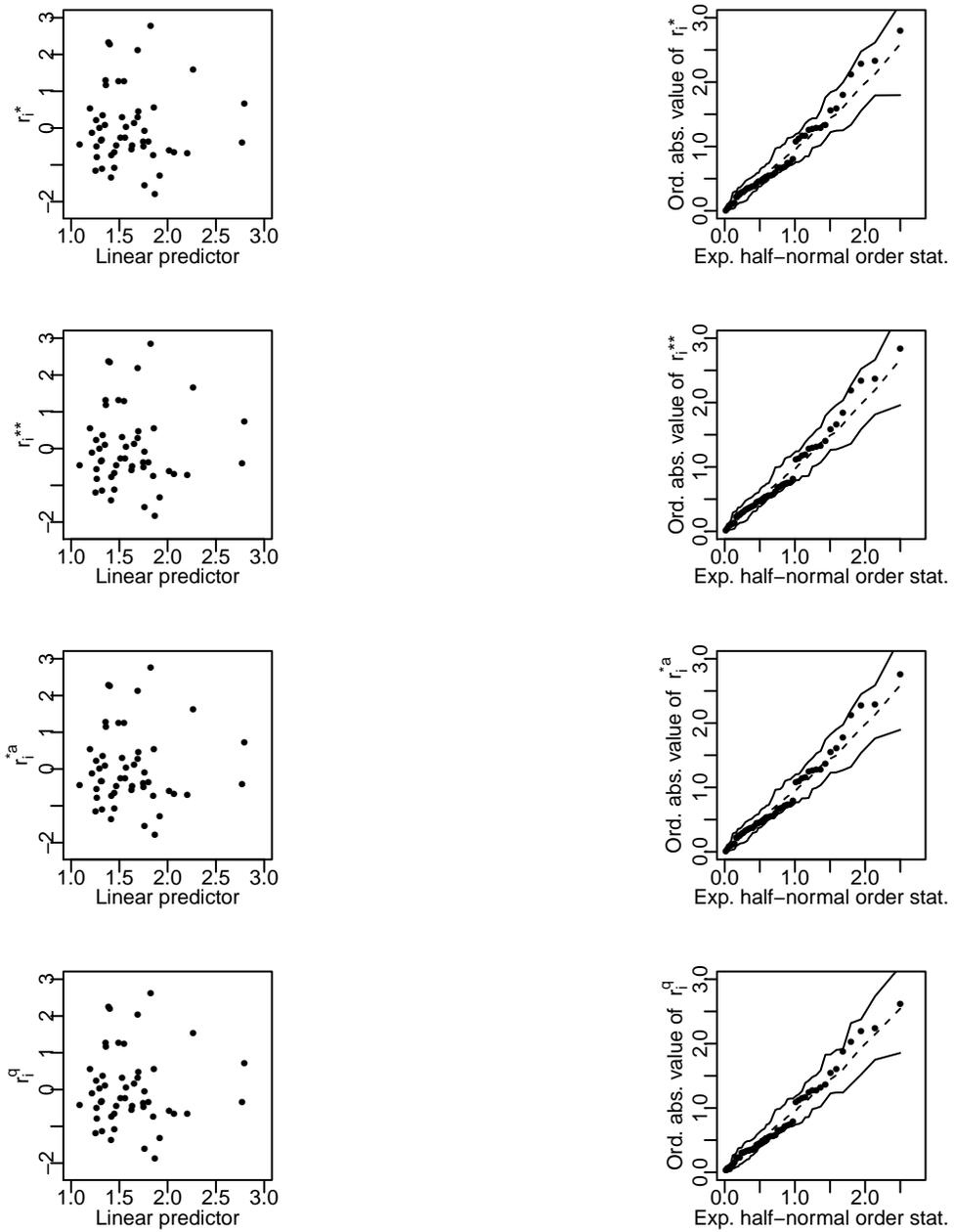}
\caption{Residuals against linear predictor (left) and half-normal residual plots with simulated envelope (right)
for tennis data.}
\label{fi:env2}
\end{figure}

The second application uses data on men professional tennis. The aim is study the proportion of service games won as a function of the number of aces per 5 games of service (AP5GS) and the first service percentage (FSP), that is, the percentage of services that the player got in on his first try. Data were obtained from ATP Website \citep{ATP94} and refer to the performance of the best 50 tennis players of the world in the 10 first months of 2015.
Figure \ref{fi:env2} presents the residuals plots of the beta regression model with logit link function.
None of the residuals suggest model misspecification. 

Table \ref{ta:aplic2} presents parameter estimates and standard errors. When the logit link function is used in beta regression, the exponential of the parameters are odds ratios \citep{fer+cri04}. It is estimated that for every ace per 5 games increase, FSP held constant, the odds of the proportion of service games won increase by 20.6\%. It is also estimated that for every percentage point increase in the first service percentage, AP5GS held constant, the odds of the proportion of service games won increase by 4.2\%. The pseudo R$^2$ of the fitted regression model is 0.719, suggesting that a substantial proportion of the variability of the proportion of service games won can be explained by a beta regression model with AP5GS and FSP as covariates.  

     \begin{table} [!hpb]
    \caption {Parameter estimates and standard errors for tennis data.}
    \begin{center}
\begin{tabular} {crrr}      
    \hline
 & Estimate & Standard error & Exp(estimate) \\
   \hline
Intercept	&	$-1.421$	&  $0.470$ &	$0.241$	\\
AP5GS	&	$0.187$	&  $0.026$ &	$1.206$	\\
FSP	&	$0.041$	&  $0.008$ &	$1.042$	\\
\hline																					
    \end{tabular}
    \end{center}
    \label{ta:aplic2}
    \end{table} 
    
In the second application, none of the residuals suggest model misspecification. However, the simulation studies presented in Section \ref{sec:simulation} indicated that $r_i^\ast$, $r_i^{\ast\ast}$ and $r_i^{\ast^a}$ could suggest lack of fit in some scenarios when the beta regression model is correct. To illustrate this fact, we simulated data with 160 observations. Sample size was ten times that used in Section \ref{sec:simulation} to reduce sample variability. Covariates and response were generated as in Scenario 1 of the simulation studies.

\begin{figure}[!htbp]\centering
\includegraphics[page=3]{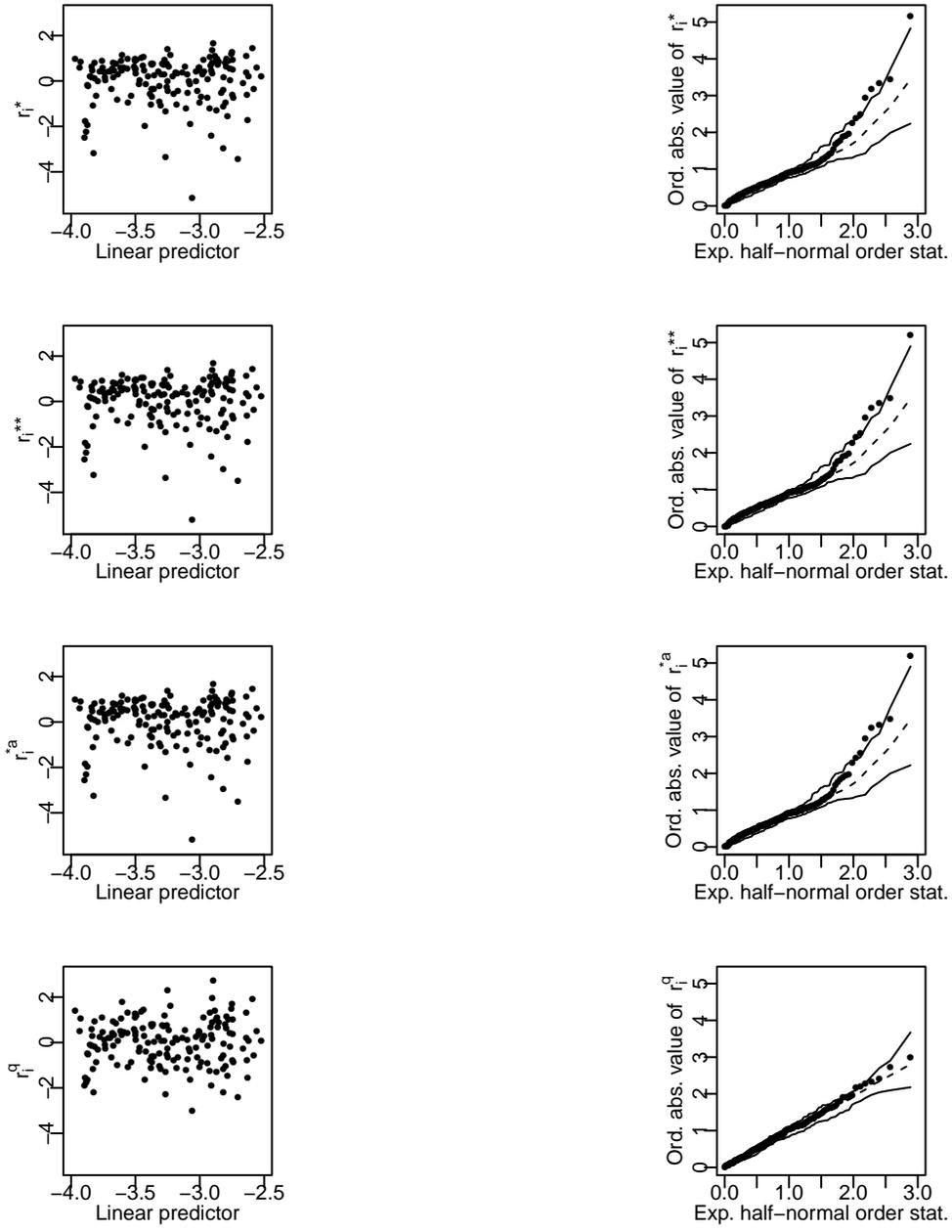}
\caption{Residuals against linear predictor (left) and half-normal residual plots with simulated envelope (right)
for simulated data.}
\label{fi:env3}
\end{figure}

Figure \ref{fi:env3} presents the residuals plots of the beta regression model with logit link function.
Residuals $r_i^\ast$, $r_i^{\ast\ast}$ and $r_i^{\ast^a}$ incorrectly suggest model misspecification because there are several residuals lower than $-3$ and four out of the five higher residuals (in absolute value) are outside the boundaries of the simulated envelope. On the other hand, plots of $r_i^q$ do not suggest lack of fit.

\section{Concluding remarks}

In this work, we studied the properties of four residuals in beta regression using three applications and Monte Carlo simulation techniques. Five scenarios were simulated for two values of the precision parameter, two sample sizes and two link functions. 

It is very desirable that a residual has the following two properties. First, it should be able to detect lack of fit of the model. Second, its distribution should be well approximated by the standard normal distribution. The first application suggested that the four residuals considered in this paper can detect lack of fit of the beta regression model. However, the simulation studies indicated that they are not similar regarding the second property. 

In all scenarios, the distribution of the weighted standardized residual 2 is worse approximated by the standard normal distribution than that of the weighted standardized residual 1. In addition, in none of the scenarios, the weighted standardized residual 1 has the closest distribution to the standard normal distribution. Therefore, these residuals do not seem to be the best choice to perform diagnostic analysis in beta regression. The best residual regarding normal approximation depends on the scenario. For $n=40$, except in the case where all mean responses are close to $0.5$, the distribution of the quantile residual is better approximated by the standard normal distribution than that of the adjusted weighted standardized residual 1. This result suggests that the quantile residual is better than the adjusted weighted standardized residual 1 for moderate to large sample sizes. When $n=16$, for most scenarios, the quantile residual is the best for $\phi=10$ and the adjusted weighted standardized residual 1 performs better for $\phi=100$. These findings seem to indicate that, for small sample sizes, the quantile residual is better than the adjusted weighted standardized residual 1 when the variance of the response variable is high, and worse in the low variance case. In addition, the distribution of the quantile residual satisfactorily approximates the standard normal distribution in all scenarios, but the distribution of the adjusted weighted standardized residual 1 displays considerable asymmetry when $\phi$ is small and the mean responses are close to one of the limits of the standard unit interval. This feature of the adjusted weighted standardized residual 1 may lead to incorrect conclusions as shown in the third application. 

Besides the advantages shown in the simulation results, the quantile residual has two other advantages compared to the adjusted weighted standardized residual 1: it is simpler and it can be used for any link function. Overall, the quantile residual is therefore a better choice to perform diagnostic analysis in beta regression than the competing residuals.

\section{Acknowledgments}

The author thanks professor Rafael Izbicki for helpful comments and suggestions on an earlier draft and Fapesp for financial support.


\singlespacing   

\bibliographystyle{elsarticle-harv}

\bibliography{bibquant_v02}  
\end{document}